\documentclass[12pt]{article}

\usepackage[document]{ragged2e}
\usepackage{amssymb}
\usepackage{amsmath} 
\usepackage{amsthm}
\usepackage{bm}
\usepackage{lipsum}
\usepackage{graphicx}
\begin{document}

\begin{titlepage}
\thispagestyle{plain}
\begin{center}
\textbf{Design Framework for Spherical Microphone and Loudspeaker Arrays in a Multiple-input Multiple-output System} \\ \vspace{10ex}
Hai Morgenstern$^{1,}$\footnote{e-mail: haimorg@post.bgu.ac.il}, Boaz Rafaely$^1$, and Markus Noisternig$^2$\\
$^{1}$Department of Electrical and Computer Engineering, \\
Ben Gurion University of the Negev, Beer- Sheva, 84105, Israel\\
$^{2}$ Acoustic and Cognitive Spaces Group, \\
IRCAM, CNRS, Sorbonne University, UPMC Paris 6, UMR 9912, STMS, \\
75004 Paris, France \\
\vspace{10ex}
Running title: Design framework for spherical MIMO system\\
\end{center}
\end{titlepage}
		
\begin{abstract}
\thispagestyle{plain}
\setcounter{page}{2}
\justify
Spherical microphone arrays (SMAs) and spherical loudspeaker arrays (SLAs) facilitate the study of room acoustics due to the three-dimensional analysis they provide.
More recently, systems that combine both arrays, referred to as multiple-input multiple-output (MIMO) systems, have been proposed due to the added spatial diversity they facilitate. 
The literature provides frameworks for designing SMAs and SLAs separately, including error analysis from which the operating frequency range of an array is defined. 
However, such a framework does not exist for the joint design of an SMA and an SLA that comprise a MIMO system.
This paper develops a design framework for MIMO systems based on a model that addresses errors and highlights the importance of a matched design.
Expanding on a free-field assumption, errors are incorporated separately for each array and error bounds are defined, facilitating error analysis for the system.  
The dependency of the error bounds on the SLA and SMA parameters is studied and it is recommended that parameters should be chosen to assure matched operating frequency ranges of the arrays in MIMO system design.
A design example is provided, demonstrating the superiority of a matched system over an unmatched system in the synthesis of directional room impulse responses.
\vspace{10ex}
				
\end{abstract}
\addtocounter{page}{3}

\section{INTRODUCTION}
\justify
\setlength{\parindent}{5ex}

Acoustic systems that employ an array (either a microphone array or a loudspeaker array) have been shown to be advantageous in a variety of applications compared to systems that employ a single microphone or a single loudspeaker.  
In room acoustics, spherical microphone arrays (SMAs) have been recently proposed for measuring directional room impulse responses (RIRs) \cite{meyer2002highly,  gover2002microphone, rafaely2004plane, park2005sound, balmages2004room,carpentier2013parametric}. 
These responses facilitate a 3-D analysis of the sound field surrounding an array and the computation of spatial room-acoustics measures, such as the directivity index (DI) and directional diffusion.
The latter make it possible to study enclosures with reference to human auditory perception. 
Similarly, loudspeaker arrays and directional loudspeakers have been proposed for studying room acoustics, due to their ability to radiate acoustic energy to selected directions in space \cite{tervo2009acoustic, pueo2004precise, poletti2005three}. 
This ability enables the excitation of individual reflection paths and improves the separability of reflections when analyzing an RIR, compared to an omnidirectional loudspeaker.
In particular, spherical loudspeaker arrays (SLAs) use a compact arrangement of loudspeakers mounted on a sphere to produce complex radiation patterns in 3-D space \cite{pasqual2010application,rafaely2011optimal,zotter2009analysis}.
Employing a compact SLA with variable directivity has also been shown to lead to perceivable differences in the sound field, also affecting source localization \cite{kazuna2016}. 
In general, the performance of spherical arrays, either an SMA or an SLA, is limited by errors introduced by the following factors \cite{rafaely2005analysis,kassakian2004characterization}.  
The use of a finite number of elements in each array results in spatial-aliasing errors, which impose limitations at high frequencies. 
Also, inaccuracies in the modeling of the arrays and in the positioning of the array elements introduce errors at all frequencies.  
While these errors are typically larger at high frequencies (due to the shorter wavelength), they also become dominant at low frequencies due to the employment of various methods for enhancing spatial resolution, such as plane-wave decomposition. 
These methods introduce ill-conditioning at low frequencies \cite{rafaely2004plane}, imposing performance limitations at these frequencies \cite{abhayapala2002theory, rafaely2007spatial,rafaely2005analysis,ward2001reproduction}. 
An analysis of all errors can be used for defining an operating frequency range (OFR) of an array, i.e., defining the operating range of frequencies in which an array performs well without introducing significant errors. 
This analysis facilitates the design of SLAs and SMAs separately, and the design and implementation of beamforming methods using these arrays.  

Until recently, SLAs and SMAs were mainly used independently. 
Systems that include both, referred to as acoustic multiple-input multiple-output (MIMO) systems, are relatively new, and the added spatial diversity that these systems provide could potentially improve the spatial analysis of room acoustics compared to systems with a single array. 
MIMO systems have been previously proposed for room acoustics analysis and three-dimensional auralization of concert halls \cite{gerzon1975recording, farina2003recording,farina2006room}. 
Although these works motivate the use of acoustic MIMO systems, they neither present a theoretical study of MIMO systems, nor of their potential performance. 
An elaborate theoretical formulation of MIMO systems was recently presented, including an example for the use of such a system in spatial analysis based on RIRs \cite{morgenstern2015theory,morgenstern2015spatial}.  
Moreover, additional examples of methods for room acoustic analysis that use MIMO systems were presented in recent conference papers \cite{morgenstern2012joint, morgenstern2013enhanced}.   
However, previous works do not provide an integrated framework for designing the arrays which constitute a MIMO system. 

In this paper it is shown that a MIMO system may become unusable due to mismatch in the OFRs of the SLA and SMA.  
To overcome this problem, a framework for the design of MIMO systems is developed, based on a model that includes errors from both of the arrays. 
Initially, an error-free model is presented in sec.~\ref{sec:System model I}. 
The model is extended, in sec.~\ref{sec:System model II}, to account for errors due to sampling at the arrays and for modeling errors. 
Error bounds are then derived in sec.~\ref{sec:errorbounds}, which are then used in sec.~\ref{sec:jointdesign} for defining the system and the arrays' OFRs in a design example.  
An analysis of these bounds shows that the arrays' OFRs should be matched to achieve a useful OFR for the MIMO system.
A method is proposed in sec.~\ref{sec:restrictedspatialresolution} for extending the OFR of a MIMO system. 
The superiority of a matched system over an unmatched system in the generation of directional RIRs is demonstrated in a simulation study in sec.~\ref{sec:desexamples}. 
Lastly, conclusions are outlined in sec.~\ref{sec:conclusions}.
\medskip

\section{System model}
\label{sec:System model I}
\justify
A model of an acoustic MIMO system in free-field is presented in this section.
For further details regarding the model formulation, a study of its properties, and its extension for systems positioned in enclosures, the reader is referred to \cite{morgenstern2015theory}. 

\begin{figure}[t]
\centering
\includegraphics[clip, trim = {5mm 90mm 70mm 40mm}, width = 1\linewidth]{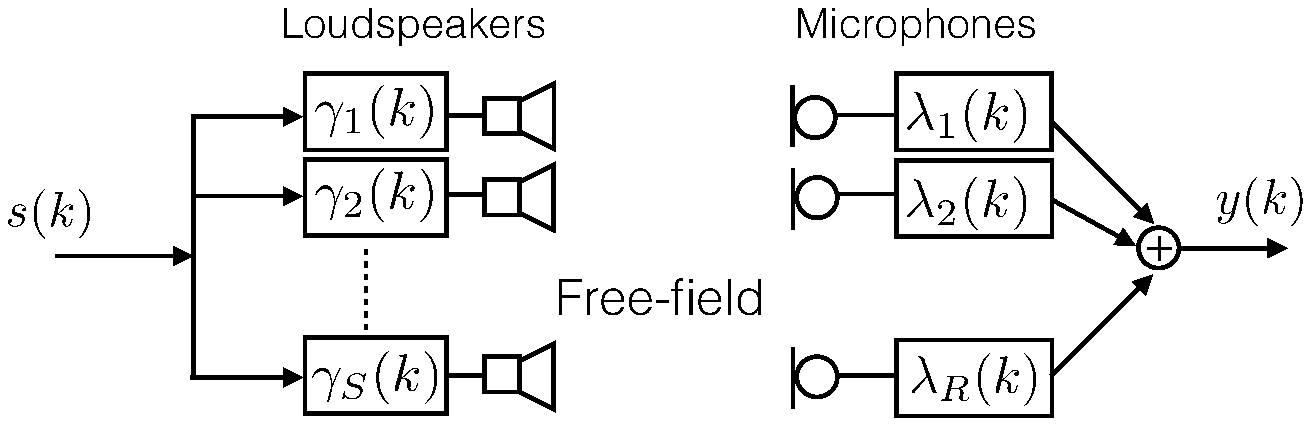}
\caption[System block diagram with SLA beamforming coefficients, $\gamma_1(k), ..,\gamma_S(k)$, and SMA beamforming coefficients, $\lambda_1(k),..., \lambda_R(k)$.]{System block diagram with SLA beamforming coefficients, $\gamma_1(k), ..,\gamma_S(k)$, and SMA beamforming coefficients, $\lambda_1(k),..., \lambda_R(k)$.}
\label{fig:1}
\end{figure}

A MIMO system comprises an SLA with $S$ loudspeakers mounted around a rigid sphere with a radius of $r_L$ and an SMA with $R$ microphones distributed over a rigid sphere with a radius of $r_M$.
Beamforming is applied to both arrays with the aim of controlling the arrays’ directivity patterns. 
At the SLA, this is achieved by weighting the SLA input signal, $s(k)$, where $k$ is the wavenumber, with complex beamforming coefficients $\gamma_1(k), ..,\gamma_S(k)$ before driving the loudspeakers. 
At the SMA, the microphones signals are weighted with complex beamforming coefficients $\lambda_1(k), ...,\lambda_R(k)$ and then summed to produce the system output, $y(k)$. 
This is formulated as:
\begin{eqnarray}\label{eq:1}
y(k) &=& \tilde{\bm \gamma}^\mathrm{T}(k) \tilde{\bm H}(k)\tilde{ \bm \lambda}(k) \,s(k), 
\end{eqnarray}
where
\begin{eqnarray}\label{eq:2}
\tilde{\bm \gamma}(k) &=& [\gamma_1(k), \gamma_2(k), ..., \gamma_S(k)]^\mathrm{T}, \\
 \tilde{\bm \lambda}(k) &=& [\lambda_1(k), \lambda_2(k), ..., \lambda_R(k)]^\mathrm{T}, \label{eq:3}
\end{eqnarray}
and $(\cdot)^{\mathrm{T}}$ is the transpose operator. 
$\tilde{\bm H}(k)$ is the $S\times R$ system transfer matrix whose elements hold the transfer functions for all loudspeaker and microphone combinations. 
For a system block diagram representing Eq.~\eqref{eq:1}, the reader is referred to Fig.~\ref{fig:1}.
Under free-field conditions and a far field assumption, $\tilde{\bm H}(k)$ can have a simple form \cite{morgenstern2015theory}: 
\begin{eqnarray}\label{eq:4}
\tilde{\bm H}(k) &=& \tilde{\bm h}_L(k) \tilde{\bm h}_M^\mathrm{H}(k),  
\end{eqnarray}
where $(\cdot)^{\mathrm{H}}$ is the complex transpose operator. 
$\tilde{\bm h}_L(k)$ is the $S\times 1$ SLA transfer vector from the input of all loudspeakers to the sound pressure at the center of the SMA, and $\tilde{\bm h}_M(k)$ the $R\times 1$ SMA transfer vector from the sound pressure at the center of the SMA to the output of all microphones \cite{morgenstern2015theory}. 
$\tilde{\bm h}_L(k)$ can be decomposed as follows \cite{morgenstern2015theory}: 
\begin{eqnarray}\label{eq:5}
\tilde{\bm h}_L(k) &=& {\bm Y}_L \tilde{\bm G}(k) \tilde{\bm \psi}_L(k), 
\end{eqnarray}
with matrix ${\bm Y}_L$ depending on the angles of the loudspeakers, matrix $\tilde{\bm G}(k)$ representing the radiation of the loudspeaker cap, and $\tilde{\bm \psi}_L(k)$ relating the centers of the SLA and SMA.
In particular, ${\bm Y}_L $ is an $S \times (\tilde{N}_L+1)^2$ matrix, given by: 
\begin{eqnarray}\label{eq:6}
{\bm Y}_L  &=& [\bm y_{\tilde{N}_L}(\bm \eta_1), \bm y_{\tilde{N}_L}(\bm \eta_2), ..., \bm y_{\tilde{N}_L}(\bm \eta_S)], 
\end{eqnarray}
with 
\begin{eqnarray}\label{eq:7}
\bm y_{\tilde{N}_L}(\bm \eta) & = & [Y_{0}^0(\bm \eta), Y_{1}^{-1}(\bm \eta), ... , Y_{\tilde{N}_L}^{\tilde{N}_L}(\bm \eta)]^{\mathrm{T}},    
\end{eqnarray}
being an $(\tilde{N}_L+1)^2 \times 1$ vector that holds spherical harmonics (SH), $\{Y_{n}^{{m}}(\bm \eta)\}$, evaluated at elevation angle $\theta$ and azimuth angle $\phi$ with respect to the SLA center, written shortly as $\bm \eta = (\theta, \phi)$. 
The SH order $\tilde{N}_L$ should tend to infinity for an exact representation of the sound field \cite{morgenstern2015theory}.
However, in practice a finite order is chosen, as will be discussed in more detail in sec.~\ref{sec:jointdesign}.
$\bm \eta_1,\bm \eta_2,..., \bm \eta_S $ are the spatial angles that point at the loudspeakers.  
\begin{eqnarray}\label{eq:8}
\tilde{\bm G}(k) &=& \text{diag}[g_{0}(kr_L), g_{1}(kr_L), ..., g_{\tilde{N}_L}(kr_L)] 
\end{eqnarray}
is an $(\tilde{N}_L+1)^2 \times (\tilde{N}_L+1)^2$ diagonal matrix that holds coefficients $g_{n}(kr_L)$ that depend on spherical Hankel and Bessel functions, the SLA array type, its radius, and the size of the loudspeaker cap \cite{morgenstern2015theory}, and
\begin{eqnarray}\label{eq:9}
\tilde{\bm \psi}_L(k) & = & \frac{e^{ikr_0}}{r_0}\bm y_{\tilde{N}_L}(\bm \eta_0),  
\end{eqnarray}
is an $(\tilde{N}_L+1)^2 \times 1$ vector, with $\bm \eta_{0}$ being the angles that point at the SMA center, where $i^2 = -1$, and $r_0$ is the distance between the centers of the arrays.  

\begin{figure}[t]
\centering
\includegraphics[clip, trim = {0 0 0 {5 mm}}, width = 1\linewidth]{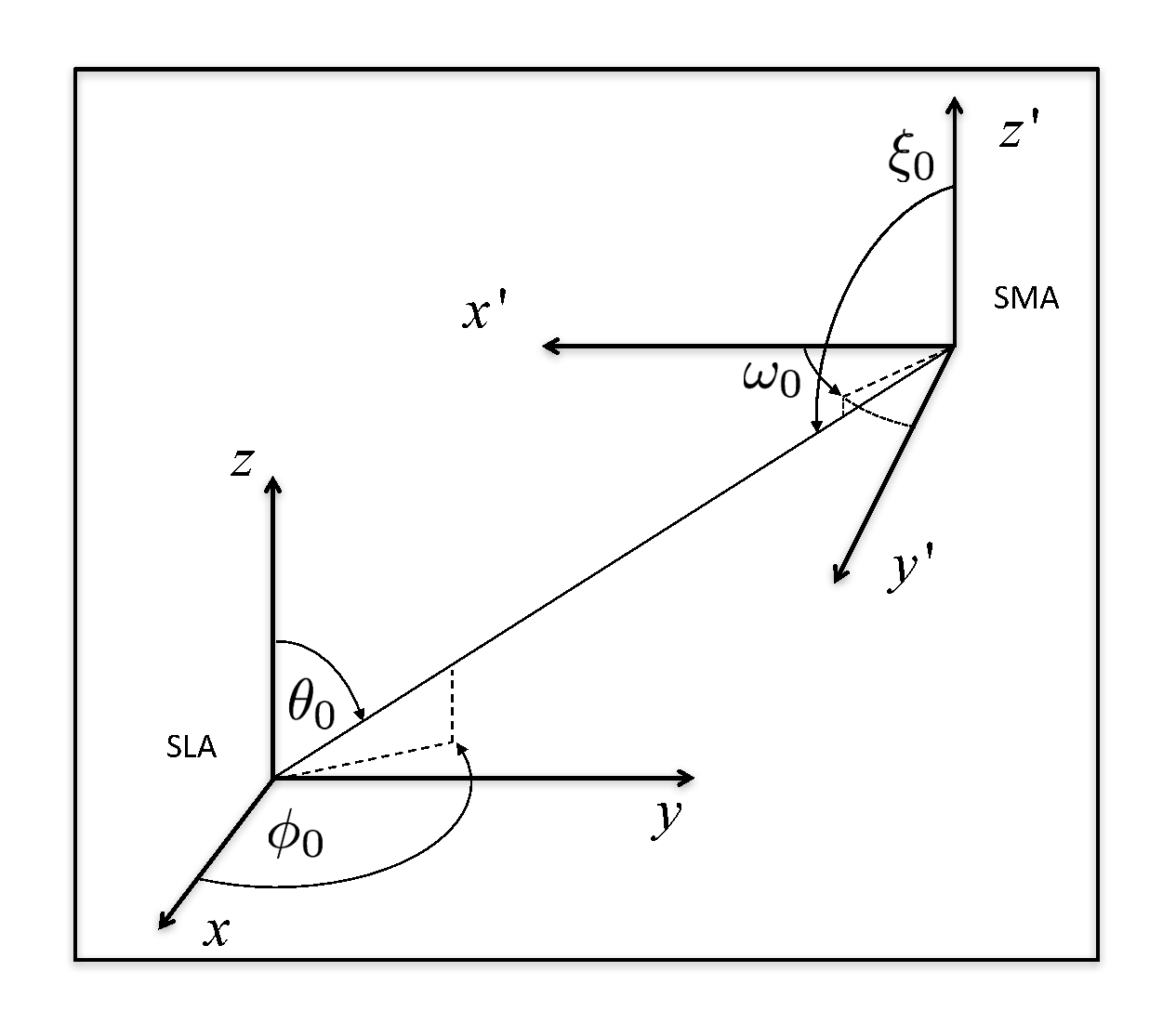}
\caption[Diagram of the SLA and the SMA, with Cartesian coordinate systems $(x,y,z)$ and $(x',y',z')$, respectively. $\bm \eta_0 =(\theta_0,\phi_0) $ and $\bm \beta_0 = (\xi_0, \omega_0)$ are the DOR and DOA, respectively.]{Diagram of the SLA and the SMA, with Cartesian coordinate systems $(x,y,z)$ and $(x',y',z')$, respectively. $\bm \eta_0 =(\theta_0,\phi_0) $ and $\bm \beta_0 = (\xi_0, \omega_0)$ are the DOR and DOA, respectively.}
\label{fig:SMA_SLA_coordinate}
\end{figure}

For the SMA, $\tilde{\bm h}_M(k)$ can be decomposed as follows:
\begin{eqnarray}\label{eq:10}
\tilde{\bm h}_M(k) &=&{\bm Y}_M^*  \tilde{\bm B}(k) \tilde{\bm \psi}_M, 
\end{eqnarray}
where $(\cdot)^*$ is the complex conjugate operator. 
${\bm Y}_M $ is an $R \times (\tilde{N}_M+1)^2$ matrix, given by: 
\begin{eqnarray}\label{eq:11}
{\bm Y}_M  &=&  [\bm y_{\tilde{N}_M}(\bm \beta_1), \bm y_{\tilde{N}_M}(\bm \beta_2), ..., \bm y_{\tilde{N}_M}(\bm \beta_R)], 
\end{eqnarray}
with $\bm y_{\tilde{N}_M}(\bm \beta)$, an $(\tilde{N}_M+1)^2 \times 1$ defined as in Eq.~\eqref{eq:7}, but for elevation and azimuth angles $\bm \beta= ( \xi, \omega)$ and SH order $\tilde{N}_M$.
$\tilde{N}_M$ should tend to infinity for an exact representation of the sound field \cite{morgenstern2015theory}. 
As in the case of the SLA, a finite order is chosen, as will be discussed in sec.~\ref{sec:jointdesign}.
$\bm \beta_1,\bm \beta_2,..., \bm \beta_R $ are the spatial angles that point at the microphones.  
\begin{eqnarray}\label{eq:12}
\tilde{\bm B}(k) &=& \text{diag}[b_{0}(kr_M), b_{1}(kr_M), ..., b_{\tilde{N}_M}(kr_M)] \nonumber \\
& &
\end{eqnarray}
is an $(\tilde{N}_M+1)^2 \times (\tilde{N}_M+1)^2 $ diagonal matrix that holds coefficients $b_{n}(kr_M)$ that depend on spherical Hankel and Bessel functions, the SMA array type, and its radius \cite{morgenstern2015theory}, and
\begin{eqnarray}\label{eq:13}
\tilde{\bm \psi}_M & =  &  \bm y_{\tilde{N}_M}(\bm \beta_0), 
\end{eqnarray}
is an $(\tilde{N}_M+1)^2\times 1$ vector, with $\bm \beta_0$ being the angles that point at the SLA center. 
$\bm \eta_{0 }$ and $\bm \beta_0$ are referred to as the direction of radiation (DOR) and direction of arrival (DOA), respectively. 
A schematic illustration of the system is presented in Fig.~\ref{fig:SMA_SLA_coordinate}. 
For ease of illustration, the coordinate system of the SMA has been rotated. 

An alternative representation of the system in the SH domain is simpler and more appropriate for spatial analysis with spherical arrays \cite{morgenstern2015theory}.
Transfer functions are formulated between $(N_L+1)^2$ SH channels of the SLA and $(N_M+1)^2$ SH channels of the SMA, instead of between the $S$ loudspeakers and the $R$ microphones, respectively. 
$N_L$, the SH order of the SLA, denotes the SH order that can be controlled with $S$ loudspeakers and, therefore, typically maintains: $(N_L+1)^2\leq S$. 
Note that $N_L$ may be different than $\tilde{N}_L$, as defined earlier (cf.~Eq.~\eqref{eq:5}).  
Similarly for the SMA, $N_M$, the SMA SH order, denotes the SH order that can be controlled with $R$ microphones, typically maintaining $(N_M+1)^2\leq R$, and may be different to $\tilde{N}_M$ (cf.~Eq.~\eqref{eq:10}). 
The system output in this case is given by applying beamforming in the SH domain as \cite{morgenstern2015theory}:
\begin{eqnarray}\label{eq:14}
y(k) &=& \bm \gamma^\mathrm{T}(k) \bm H(k) \bm \lambda(k) \,s(k), 
\end{eqnarray}
where 
\begin{eqnarray}\label{eq:15}
{\bm \gamma}(k) &=& [\gamma_{00}(k), \gamma_{1(-1)}(k), \gamma_{10}, \gamma_{11}(k), ... \nonumber \\\
& & ..., \gamma_{N_L(N_L-1)}(k), \gamma_{N_LN_L}(k)]^\mathrm{T} \, \, \text{and} \\
{\bm \lambda}(k) &=& [\lambda_{00}(k), \lambda_{1(-1)}(k), \lambda_{10}(k), \lambda_{11}(k), ... \nonumber \\
& & ... , \lambda_{N_M(N_M-1)}(k), \lambda_{N_MN_M}(k)]^\mathrm{T}, \label{eq:16}
\end{eqnarray}
are the $(N_L+1)^2\times 1$ SLA beamforming coefficients vector and the $(N_M+1)^2\times 1$ SMA beamforming coefficients vector, respectively. 
$\bm H(k)$ is an $(N_L+1)^2\times (N_M+1)^2$ matrix given by: 
\begin{eqnarray}\label{eq:17}
\bm H(k) & = & \bm h_L(k) \bm h_M^\mathrm{H}(k),  
\end{eqnarray}
where $\bm h_L(k)$ and $\bm h_M(k)$ are the $(N_L+1)^2 \times 1$ SLA and $(N_M+1)^2 \times 1$ SMA transfer vectors, respectively, represented in SH.  
$\bm h_L(k)$ is given as: 
\begin{eqnarray}\label{eq:18}
\bm h_L(k) &=& \bm G(k) \bm \psi_L(k), 
\end{eqnarray}
with $\bm G(k)$ and $\bm \psi_L(k)$ defined as in Eqs.~\eqref{eq:8} and \eqref{eq:9}, respectively, but for SH order $N_L$ instead of $\tilde{N}_L$. 
For the SMA, $\bm h_M(k)$ is given by: 
\begin{eqnarray}\label{eq:19}
\bm h_M(k) &=&\bm B(k) \bm \psi_M,
\end{eqnarray}
with $\bm B(k)$ and $\bm \psi_M$ defined as in Eqs.~\eqref{eq:12} and \eqref{eq:13}, respectively, but for SH order $N_M$ instead of $\tilde{N}_M$. 
\medskip

\section{Model extension - consideration of errors}\label{sec:System model II}
\justify
The system model presented in the last section and in ref.~\cite{morgenstern2015theory} is useful;
however, it does not include the effects of spatial sampling and model mismatch errors, which is necessary for this paper. 
Therefore, the model is now extended such that it includes these errors.
In particular, since $\tilde{\bm H}(k)$ from Eq.~\eqref{eq:4} is represented as an outer product of the SLA and SMA transfer vectors, the errors are added to each array separately. 
First, model mismatch errors are incorporated in $\tilde{\bm h}_L(k)$ and $\tilde{\bm h}_M(k)$ in sec.~\ref{sec:mismatch errors}. 
Then, errors caused by spatial sampling at the arrays are incorporated in secs.~\ref{sec:spatial aliasing} and \ref{sec:spurious_harmonics} for the SMA and SLA, respectively.   
Finally, in sec.~\ref{sec:normalization}, normalization is applied to both of the arrays, in an attempt to enhance their overall spatial resolution. 
\medskip

\subsection{Model mismatch}\label{sec:mismatch errors}
\justify
Model-mismatch errors are caused by several factors. 
These include uncertainties in the positions and frequency responses of loudspeakers and microphones, inaccuracies in the model represented by the values of $g_{n}(kr_L)$ and $b_{n}(kr_M)$ from Eqs.~\eqref{eq:8} and \eqref{eq:12}, respectively (which may arise, for example, when the assumption of a rigid sphere is violated for either the SLA or the SMA), and errors caused by computation with finite numerical precision (such as in sound cards). 
While such errors can be modeled accurately, given information on the sources of the errors, a generalized additive error model is employed in this paper. 
Such a model facilitates a matched design of a MIMO system that is independent of the exact model of the errors, since these may vary for different conditions and systems. 
Moreover, since the effect of model-mismatch errors has been shown to be similar to the effect of transducer noise \cite{rafaely2005analysis}, this general error model also facilitates a study of the system robustness against transducer noise. 

Error vectors are added to the SMA and SLA transfer vectors, represented in the spatial domain. 
At the SLA, the SLA transfer vector with error is written, for simplicity, using the same notation as the SLA transfer vector without error.
$\tilde{\bm h}_L(k)$ from Eq.~\eqref{eq:5} is therefore updated to: 
\begin{eqnarray}\label{eq:25}
\tilde{\bm h}_L(k) &=& {\bm Y}_L \tilde{\bm G}(k)\tilde{\bm \psi}_L(k) + \tilde{\bm n}_L(k), 
\end{eqnarray}
where $\tilde{\bm n}_L(k)$ is the $S \times 1$ SLA error vector, given by: 
\begin{eqnarray}\label{eq:26}
\tilde{\bm n}_L(k) &=& [n^L_1(k), n^L_2(k), ...., n^L_S(k)]^\mathrm{T},   
\end{eqnarray}
in which $n^L_1(k),n^L_2(k),...n^L_S(k)$ are assumed to be independent and identically distributed (i.i.d.). 
In particular, the variances of these errors can be set with respect to the average power of the loudspeaker transfer functions; 
this will be discussed in sec.~\ref{sec:jointdesign}.
Similarly for the SMA, the notation of $\tilde{\bm h}_M(k)$ is kept for simplicity and Eq.~\eqref{eq:10} is updated to: 
\begin{eqnarray}\label{eq:27}
\tilde{\bm h}_M(k) &=& {\bm Y}_M \tilde{\bm B}(k) \tilde{\bm \psi}_M + \tilde{\bm n}_M(k), 
\end{eqnarray}
where $\tilde{\bm n}_M(k)$ is the $R \times 1$ SMA error vector given by: 
\begin{eqnarray}\label{eq:28}
\tilde{\bm n}_M(k) &=& [n^M_1(k), n^M_2(k), ...., n^M_R(k)]^\mathrm{T},  
\end{eqnarray}
in which $n^M_1(k),n^M_2(k),...n^M_R(k)$ are, as in the case of the SLA, assumed to be i.i.d.
\medskip

\subsection{Spatial aliasing at the SMA}\label{sec:spatial aliasing}
\justify
Spatial sampling may introduce aliasing errors that are non-negligible at high frequencies \cite{rafaely2005analysis}.
For microphone arrays, spatial sampling is the process in which the microphones that comprise an array sample the sound field at the array surface. 
In particular for SMAs, an important feature of sampling is to facilitate the computation of the spherical Fourier transform (SFT) of the sound field on the array's surface, which gives its SH representation. 
The computation of the SFT is performed by applying sampling weights to the SMA elements \cite{meyer2002highly,rafaely2005analysis,li2007flexible}.
Applying the weights on $\tilde{\bm h}_M(k)$ from Eq.~\eqref{eq:27}, a new SMA transfer vector $\hat{\bm h}_M(k)$, represented in the SH domain, can be defined be as: 
\begin{eqnarray}\label{eq:aliasing1}
\hat{\bm h}_M(k) & = &  \bm \alpha_M  \left[  {\bm Y}_M \tilde{\bm B}(k) \tilde{\bm \psi}_M + \tilde{\bm n}_M(k) \right], 
\end{eqnarray}
where 
\begin{eqnarray}\label{eq:bounds30}
\bm \alpha_M & = &  \begin{bmatrix}  \alpha_{1,00}^M & ... & \alpha_{R,00}^M \\  \vdots & \ddots & \vdots \\   \alpha_{1,N_MN_M}^M & ... & \alpha_{R,N_MN_M}^M \end{bmatrix}
\end{eqnarray}
is an $(N_M+1)^2 \times R$ matrix that holds the SMA sampling weights, leading to $\hat{\bm h}_M(k)$ that is of dimensions $(N_M+1)^2 \times 1$.
When sampling weights that are ideal for SH order $N_M$ are chosen, the following identity holds \cite{rafaely2005analysis}: 
\begin{eqnarray}\label{eq:31}
\bm \alpha_M {\bm Y}_M    & = & [\bm I_M \,\,\bm \epsilon_M], 
\end{eqnarray}
where $\bm I_M$ is the $(N_M+1)^2 \times (N_M+1)^2$ identity matrix and $\bm \epsilon_M$ is an $(N_M+1)^2 \times [(\tilde{N}_M+1)^2- (N_M+1)^2]$ matrix that holds the SMA aliasing function up to order $\tilde{N}_M$ \cite{rafaely2005analysis}. 
Using this identity and the definition from Eq.~\eqref{eq:19}, Eq.~\eqref{eq:aliasing1} can be rewritten as: 
\begin{eqnarray}\label{eq:aliasing2}
\hat{\bm h}_M(k)   & = &   
[\bm I_M \,\,\bm \epsilon_M]\tilde{\bm B}(k)\tilde{\bm \psi}_M +  \bm \alpha_M\tilde{\bm n}_M(k) \nonumber \\
& = & [\bm I_M \,\,{\bm 0}_M]\tilde{\bm B}(k)\tilde{\bm \psi}_M + [\tilde{\bm 0}_M \,\,\bm \epsilon_M]\tilde{\bm B}(k)\tilde{\bm \psi}_M +  \nonumber \\
& & +\bm \alpha_M\tilde{\bm n}_M(k) \nonumber \\
& =  & \bm B(k)\bm \psi_M  + \bm \epsilon_M \tilde{\bm z}_M(k)+  \bm \alpha_M\tilde{\bm n}_M(k), 
\end{eqnarray}
where ${\bm 0}_M$ and $\tilde{\bm 0}_M $ are matrices of zeros with the same dimensions as those of $\bm \epsilon_M$ and $\bm I_M$, respectively, and $\tilde{\bm z}_M(k)$ is defined:
\begin{eqnarray}\label{eq:aliasing3}
\tilde{\bm z}_M & = & [{\bm 0}_M^{\mathrm{H}} \, \, \tilde{\bm I}_M  ]  \tilde{\bm B}(k)\tilde{\bm \psi}_M, 
\end{eqnarray}
with $\tilde{\bm I}_M$ being the $[(\tilde{N}_M+1)^2 - (N_M+1)^2] \times [(\tilde{N}_M+1)^2 - (N_M+1)^2]$ identity matrix.
$\tilde{\bm z}_M(k)$, therefore, represents coefficients of the SMA transfer vector for orders higher than $N_M$. 
These orders are projected into the first $(N_M+1)^2$ orders via multiplication with the SMA aliasing coefficients in $\bm \epsilon_M$. 
This is referred to as the spatial aliasing of orders higher than $N_M$.
Finally, note that after applying the sampling weights, beamforming will be applied to $\hat{\bm h}_M(k)$ in Eq.~\eqref{eq:aliasing1} using $\bm \lambda$, and not $\tilde{\bm \lambda}$. 
In practice, sampling weights and beamforming are sometimes applied in a single stage, by applying beamforming vector $\tilde{\bm \lambda} =  \bm \alpha_M^{H}{\bm \lambda}$ directly to $\tilde{\bm h}_M(k)$ from Eq.~\eqref{eq:27}. 
\medskip

\subsection{Spurious harmonics at the SLA}\label{sec:spurious_harmonics}
\justify
Spatial sampling by loudspeakers may introduce errors that are non-negligible at high frequencies, as in the case of SMAs.
Applying sampling weights to the loudspeakers that comprise an SLA has been proposed in ref.~\cite{atkins2010optimal}, in order to calculate beamforming coefficients that produce a desired sound field represented using SH functions. 
In this work, sampling at the SLA facilitates a representation of the transfer vectors from the SLA input in the SH domain to the sound pressure at the center of the SMA (instead of from loudspeakers to the sound pressure at the center of the SMA). 
Given $\tilde{\bm h}_L(k) $ from Eq.~\eqref{eq:25}, a new SLA transfer vector $\hat{\bm h}_L(k)$ can be defined as: 
\begin{eqnarray}\label{eq:spurious1}
\hat{\bm h}_L(k) &=& \bm \alpha_L \left[ {\bm Y}_L \tilde{\bm G}(k) \tilde{\bm \psi}_L(k) + \tilde{\bm n}_L(k) \right], 
\end{eqnarray}
where 
\begin{eqnarray}\label{eq:35}
\bm \alpha_L & = &  \begin{bmatrix}  \alpha_{1,00}^L & ... & \alpha_{S,00}^L \\  \vdots & \ddots & \vdots \\   \alpha_{1,N_LN_L}^L & ... & \alpha_{S,N_LN_L}^L \end{bmatrix}
\end{eqnarray}
is an $(N_L+1)^2 \times S$ matrix that holds the SLA sampling weights.
Similar to the SMA, when sampling weights are chosen that are ideal for order $N_L$, the following identity holds \cite{rafaely2005analysis}: 
\begin{eqnarray}\label{eq:36}
\bm \alpha_L {\bm Y}_L    & = & [\bm I_L \,\,\bm \epsilon_L], 
\end{eqnarray}
where $\bm I_L$ is an $(N_L+1)^2 \times (N_L+1)^2$ identity matrix and $\bm \epsilon_L$ is an $(N_L+1)^2 \times [(\tilde{N}_L+1)^2- (N_L+1)^2]$ matrix that holds the SLA spatial aliasing function up to order $\tilde{N}_L$.  
Using this identity and the definition in Eq.~\eqref{eq:18}, Eq.~\eqref{eq:spurious1} can be rewritten as: 
\begin{eqnarray}\label{eq:spurious2}
\hat{\bm h}_L(k) &=&   [\bm I_L \,\,\bm \epsilon_L] \tilde{\bm G}(k) \tilde{\bm \psi}_L(k) + \bm \alpha_L\tilde{\bm n}_L(k) \nonumber \\
& = & [\bm I_L \,\,{\bm 0}_L]\tilde{\bm G}(k)\tilde{\bm \psi}_L + [\tilde{\bm 0}_L \,\,\bm \epsilon_L]\tilde{\bm G}(k)\tilde{\bm \psi}_L +  \nonumber \\
& & +\bm \alpha_L\tilde{\bm n}_L(k) \nonumber \\
& = & \bm G(k) \bm \psi_L(k) + \bm \epsilon_L \tilde{\bm z}_L(k)  + \bm \alpha_L\tilde{\bm n}_L(k), 
\end{eqnarray}
where ${\bm 0}_L$ and $\tilde{\bm 0}_L$ are matrices of zeros with the same dimensions as those of $\bm \epsilon_L$ and $\bm I_L$, respectively, and $\tilde{\bm z}_L(k)$ is defined:
\begin{eqnarray}\label{eq:spurious3} 
\tilde{\bm z}_L & = & [{\bm 0}_L^{\mathrm{H}} \, \,  \tilde{\bm I}_L  ]  \tilde{\bm G}(k)\tilde{\psi}_L, 
\end{eqnarray}
with $\tilde{\bm I}_L$ being the $[(\tilde{N}_L+1)^2 - (N_L+1)^2] \times [(\tilde{N}_L+1)^2 - (N_L+1)^2]$ identity matrix. 
$\tilde{\bm z}_L(k)$, therefore, represents coefficients of the SLA transfer vector for orders higher than $N_L$. 
We refer to these coefficients as spurious, or artificial, harmonics, and they can also be viewed as the equivalent of spatial aliasing in SMAs.
While spatial aliasing in SMAs refers to the aliasing of higher-order SH coefficients into low-order SH coefficients due to sampling, in SLAs, higher order SH are generated due to the sampling of a low-order function on a sphere.
Low-order harmonics of a continuous function on a sphere are projected to these spurious, higher-order harmonics via multiplication with $\bm \epsilon_L$. 
Due to the resemblance of the spurious harmonics to spatial aliasing, results from the SMA literature can be directly applied to SLAs. 
For example, spatial aliasing is reduced by the SMA through the attenuation of $b_{n}(kr_M)$ from Eq.~\eqref{eq:12} for $n>kr_M$ \cite{rafaely2005analysis}.
In a similar manner, spurious harmonics are also reduced when radiated to the far field through the attenuation of  $g_{n}(kr_L)$ from Eq.~\eqref{eq:8}, for $n>kr_L$.
Similarly, methods proposed for minimizing spatial aliasing, such as those described in refs.~\cite{rafaely2007spatial,alon2012spherical}, can potentially be adapted and employed for minimizing the energy of spurious harmonics.
Finally, note that after applying the sampling weights, beamforming will be applied to $\hat{\bm h}_L(k)$ in Eq.~\eqref{eq:spurious2} using ${\bm \gamma}$, and not $\tilde{\bm \gamma}$. 
Similarly to SMAs, sampling weights and beamforming are sometimes applied in a single stage, by applying $\tilde{\bm \gamma}  = \bm \alpha_L^{H}{\bm \gamma}$ directly to $\tilde{\bm h}_L(k)$ from Eq.~\eqref{eq:25}. 
In particular, this approach for formulating $\tilde{\bm \gamma}$ yields the same beamforming vector as in other methods proposed in the literature for SLAs \cite{wu2009theory,poletti2005three,daniel2003further,zotter2007modeling}.
\medskip

\subsection{System normalization}\label{sec:normalization}
\justify
Normalization can be applied in order to allow for a description of the measured sound field that is independent of the arrays’ physical construction, facilitating a representation of that sound field using plane waves \cite{rafaely2004plane}.
In particular, it was shown in ref.~\cite{morgenstern2015theory} that normalization leads to improved performance in room acoustics applications.
We consider it to be important to incorporate normalization when developing a framework for a joint design of the arrays in a MIMO system;
normalization is widely used in spherical array processing, but its employment introduces ill-conditioning at low frequencies.
Hence, its investigation at the system level is important. 

Normalization is applied to the arrays of the MIMO system;
at the SLA the SH channels are normalized by the functions $g_n(kr_L)$ from Eq.~\eqref{eq:8}, while at the SMA, SH channels are normalized by $b_n(kr_M)$ from Eq.~\eqref{eq:12}.  
In practice, this is performed together with beamforming, by applying a normalized beamforming vector $\bm v(k)  =  [\bm G(k)^{-1}]^{\mathrm{H}} \bm  \gamma(k)$ to $\hat{\bm h}_L(k)$ from Eq.~\eqref{eq:spurious2} for the SLA, and by applying vector $\bm u(k)  = [\bm B(k)^{-1}]^{\mathrm{H}}  \bm \lambda(k)$ from Eq.~\eqref{eq:aliasing2} for the SMA. 
In this work, normalization is formulated inherently in the system model for deriving error bounds that are beamforming-independent in the following section.
Therefore, normalized system transfer vectors and matrix are defined. 
For the SLA, an $(N_L+1)^2 \times 1$ normalized transfer vector is defined using $\hat{\bm h}_L(k)$ from Eq.~\eqref{eq:spurious2} as:
\begin{eqnarray}\label{eq:45}
\hat{\bm \psi}_L(k) & = &  \bm G(k)^{-1} \hat{\bm h}_L(k) \nonumber \\
& = &   \bm \psi_L(k)  + \bm z_L(k) +\bm n_L(k), 
\end{eqnarray}
where $\bm \psi_L(k)$ is the error-free transfer vector given in Eq.~\eqref{eq:18}, 
\begin{eqnarray}\label{eq:47}
{\bm z}_L(k) & = & \bm G(k)^{-1} \bm \epsilon_L \tilde{\bm z}_L(k),\,\, \text{and} \\
{\bm n}_L(k) & = & \bm G(k)^{-1} \bm \alpha_L \tilde{\bm n}_L(k). \label{eq:48}
\end{eqnarray}
Similarly, for the SMA, an $(N_M+1)^2 \times 1$ normalized transfer vector is defined using $\hat{\bm h}_M(k)$ from Eq.~\eqref{eq:aliasing2} as:
\begin{eqnarray}\label{eq:46}
\hat{\bm \psi}_M(k) & = &   \bm B(k)^{-1} \hat{\bm h}_M(k) \nonumber \\
& = & \bm \psi_M  + \bm z_M(k) + \bm n_M(k), 
\end{eqnarray}
where $\bm \psi_M$ is the error-free transfer vector given in Eq.~\eqref{eq:19}, 
\begin{eqnarray}
{\bm z}_M(k) & = & \bm B(k)^{-1} \bm \epsilon_M\tilde{\bm z}_M(k),\,\, \text{and} \label{eq:49}\\
{\bm n}_M(k) & = & \bm B(k)^{-1} \bm \alpha_M \tilde{\bm n}_M(k).\label{eq:50}
\end{eqnarray}
Given $\hat{\bm \psi}_L(k)$ and $\hat{\bm \psi}_M(k)$, an $(N_L+1)^2 \times (N_M+1)^2$ normalized transfer matrix of the system is defined as: 
\begin{eqnarray}\label{eq:44}
\hat{\bm \Psi}(k) & = &    \hat{\bm \psi}_L(k) \hat{\bm \psi}^\mathrm{H}_M(k). 
\end{eqnarray}
Finally, an $(N_L+1)^2 \times (N_M+1)^2$ error-free normalized transfer matrix is defined as:
\begin{eqnarray}\label{eq:42}
\bm \Psi(k)  & = & \bm \psi_L(k) \bm \psi_M^\mathrm{H},
\end{eqnarray} 
facilitating the formulation of errors in the next section. 
\medskip

\section{Error formulation}\label{sec:errorbounds}
\justify
Expressions for the errors and error bounds are formulated in this section, based on the MIMO transfer matrices defined in the previous section. 
Employing a matrix-based formulation, errors and error bounds that are independent of beamforming at the arrays are derived.  

The total system error $\delta(k)$ is defined using matrices $\hat{\bm \Psi}(k)$ and ${\bm \Psi}(k)$ from Eqs.~\eqref{eq:44} and \eqref{eq:42}, respectively, as:
\begin{eqnarray}\label{eq:51}
\delta(k) & = & \frac{||{\bm \Psi}(k)- \hat{\bm \Psi}(k)||}{{||\bm \Psi}(k)||}, 
\end{eqnarray}
where $||(\cdot) ||$ denotes the standard 2-norm. 
Substituting $\hat{\bm \Psi}(k)$ and ${\bm \Psi}(k)$
and employing the Cauchy-Schwarz inequality, $\delta(k)$ can be shown to be bounded by:
\begin{eqnarray}\label{eq:52}
\delta(k)
& \leq & \frac{1}{|| \bm \psi_L(k) \bm \psi_M^\mathrm{H}||} \times \nonumber \\
& & \times \Bigg( 
 \bigg | \bigg |\bm \psi_L(k)\bm z_M^\mathrm{H}(k)\bigg | \bigg | 
+ \bigg | \bigg |\bm \psi_L(k){\bm n}_M^\mathrm{H}(k)\bigg | \bigg | +  \bigg | \bigg | {\bm z}_L(k)\bm \psi_M^\mathrm{H} \bigg | \bigg |
+ \bigg | \bigg |{\bm n}_L(k)\bm \psi_M^\mathrm{H}\bigg | \bigg | + \nonumber \\
& & +  \bigg | \bigg | {\bm z}_L(k){\bm z}_M^\mathrm{H}(k)  \bigg | \bigg | 
 +  \bigg | \bigg | {\bm z}_L(k){\bm n}_M^\mathrm{H}(k) \bigg | \bigg |  +  \bigg | \bigg | {\bm n}_L(k){\bm z}_M^\mathrm{H}(k) \bigg | \bigg | 
 +  \bigg | \bigg | {\bm n}_L(k){\bm n}_M^\mathrm{H}(k) \bigg | \bigg |  \Bigg).
 \end{eqnarray}
Note that all matrices in the last equation are outer products of two vectors. 
For such matrices, the 2-norm is equal to the product between the 2-norms of the two vectors that compose them; 
i.e., for vectors $\bm q$ and $\bm w$, the following holds: 
\begin{eqnarray}\label{eq:53}
|| \bm q \, \bm w^\mathrm{H} || &  =  & || \bm q || \, || \bm w ||,
\end{eqnarray}
which can be shown directly using the definition of the matrix norm.
Using this identity, $\delta(k)$ can be shown to be bounded by: 
\begin{eqnarray}\label{eq:54}
\delta(k) & \leq & \mathit{a}_{L}(k) + \mathit{m}_{L}(k) + 
\mathit{a}_{M}(k) + \mathit{m}_{M}(k) 
+ \left[\mathit{a}_{L}(k) + \mathit{m}_{L}(k)\right]\left[\mathit{a}_{M}(k) + \mathit{m}_{M}(k)\right],
\end{eqnarray}
where
\begin{eqnarray}\label{eq:55}
\mathit{a}_{L}(k) & = & \frac{||\bm z_L(k) ||}{||\bm \psi_L(k) ||}, 
\end{eqnarray}
\begin{eqnarray} \label{eq:56}
\mathit{m}_{L}(k) & = &  \frac{||\bm n_L(k) ||}{||\bm \psi_L(k) ||},  
\end{eqnarray}
\begin{eqnarray} \label{eq:57}
\mathit{a}_{M}(k) & = & \frac{||\bm z_M(k) ||}{|| \bm \psi_M  ||},\,\, \text{and} 
\end{eqnarray}
\begin{eqnarray} \label{eq:58}
\mathit{m}_{M}(k) & = & \frac{||\bm n_M(k) ||}{|| \bm \psi_M ||}. 
\end{eqnarray}
are referred to as the spurious-harmonics, the SLA model-mismatch, the spatial-aliasing, and the SMA model-mismatch error bounds, repsectively. 
The first two terms from Eq.~\eqref{eq:54} are referred to as the SLA error bounds since they depend only on the SLA;  
An SLA-only error can be defined as: 
\begin{eqnarray}\label{eq:59}
\delta_L(k) & = & \frac{||\bm \psi_L(k)- \hat{\bm \psi}_L(k)||}{||\bm \psi_L(k)||}.
\end{eqnarray}
By substituting $\bm \psi_L(k)$ and $\hat{\bm \psi}_L(k)$ from Eqs.~\eqref{eq:18} and \eqref{eq:45}, respectively, $\delta_L(k)$ can be shown to bounded by the sum of the SLA-related error terms from Eq.~\eqref{eq:54}:
\begin{eqnarray}\label{eq:60}
\delta_L(k) & \leq &  \mathit{a}_{L}(k) + \mathit{m}_{L}(k).  
\end{eqnarray}
Similarly, $\mathit{a}_{M}(k)$ and $\mathit{m}_{M}(k)$ are referred to as the SMA error bounds since they depend only on the SMA.
Defining an SMA-only error as: 
\begin{eqnarray}\label{eq:61}
\delta_M(k) & = & \frac{||\bm \psi_M- \hat{\bm \psi}_M(k)||}{||\bm \psi_M(k)||},
\end{eqnarray}
and substituting $\bm \psi_M(k)$ and $\hat{\bm \psi}_M(k)$ from Eqs.~\eqref{eq:19} and \eqref{eq:46}, respectively, $\delta_M(k)$ can be shown to be bounded by:
\begin{eqnarray}\label{eq:62}
\delta_M(k) & \leq &\mathit{a}_{M}(k) + \mathit{m}_{M}(k). 
\end{eqnarray}
Finally, the last four terms from the bottom line of Eq.~\eqref{eq:54} are referred to as the joint SLA-SMA errors. 
These errors are given as products of the SLA and SMA errors. 
Having defined the seperate SLA and SMA errors, $\delta(k)$ from Eq.~\eqref{eq:54} can be bounded by: 
\begin{eqnarray}\label{eq:54new}
\delta(k) & \leq & \delta_L(k) + \delta_M(k) +  \delta_L(k)\delta_M(k).
\end{eqnarray}
It is thus sufficient to bound the SLA and SMA errors in order to ensure that the energy of the total system error is bounded, justifying a system design based on the error bounds of the separate arrays.  
\medskip

\section{Matched design of an SMA and an SLA}\label{sec:jointdesign}
\justify
The error bounds defined in the previous section facilitate a general system design, in which multiple beamforming vectors can be applied to both arrays. 
The design is conducted through an error analysis for MIMO systems with different system and error parameters. 
System parameters include the array types (i.e., rigid-sphere, open-sphere, etc.), the number of elements in each array and the arrays' radii. 
The error parameters are used for generating model mismatch errors. 
Guidelines for system design and the definition of a matched MIMO system are given in this section through a design example. 

A design example is given for two MIMO systems that vary in only one of the system parameters: the SMA radius. 
The first system, \textit{SYS}1, is comprised of an SLA and an SMA with equal radii of $r_L = r_M = 0.2\,$m. 
Both arrays are rigid-sphere arrays. 
$S= 144$ loudspeakers are uniformaly distributed over the surface of the SLA, which yields an order of $N_L=8$. 
The size of the loudspeakers' cap size, which is required for calculating coefficients $g_n(kr_L)$ from Eq.~\eqref{eq:8}, is set to correspond to a 2''-diameter loudspeaker.
As for the SMA, $R=162$ microphones are distributed using a Gaussian distribution, which yields the same SH order as that of the SLA, i.e., $N_M = 8$. 
A distance of $r_0=1\,$m is set between the centers of the arrays. 
The orientations of the arrays are set such that the SMA is located at the direction of the north pole with respect to the SLA coordinate system (i.e., $(\theta_0, \phi_0) = (0 ,0)$), and vice versa, i.e., $(\xi_0, \omega_0) = (0,0)$.  
For the second system, \textit{SYS}2, the SMA radius is set to $r_M = 0.04\,$m, while all the other system parameters are the same as in \textit{SYS}1. 

The total system error, the SLA and SMA errors, and the error bounds are calculated for both systems, for frequencies between $f_{min} = 30\,$Hz and $f_{max} = 10\,$kHz. 
Array transfer and error vectors are initially generated for calculating these errors and error bounds. 
$\bm \psi_L(k)$ and $\bm \psi_M$ are generated as in Eqs.~\eqref{eq:18} and\eqref{eq:19}, respectively, given the system configuration.  
$\hat{\bm \psi}_L(k)$ and $\hat{\bm \psi}_M$ from Eqs.~\eqref{eq:45} and \eqref{eq:46}, respectively, are generated using $\bm z_L(k)$, $\bm n_L(k)$, $\bm z_M(k)$, and $\bm n_M(k)$. 
$\bm z_L(k)$ and $\bm z_M(k)$ are generated as in Eqs.~\eqref{eq:47} and \eqref{eq:49}, respectively. 
This requires choosing $\tilde{N}_L$ and $\tilde{N}_M$ for determining the dimensions of $\bm \epsilon_L$, $\tilde{\bm z}_L(k)$, $\bm \epsilon_M$, and $\tilde{\bm z}_M(k)$. 
For simplicity, a single SH order $\tilde{N} = \tilde{N}_L = \tilde{N}_M$ was used. 
$\tilde{N}$ was chosen in accordance with $f_{max}$ and the arrays' radii, based on a criterion proposed in Ref.~ \cite{rafaely2005analysis}, and was set as:
\begin{eqnarray}\label{eq:63}
\tilde{N} & = & 
\left \lceil{ \text{max}(r_M, r_L) \frac{2\pi f_{max}}{c}  }\right \rceil  + 2,
\end{eqnarray}
yielding $\tilde{N} = 39$, where $\left \lceil{\cdot }\right \rceil$ is the ceiling operator. 
This order was chosen so that the contribution of orders higher than $\tilde{N}$, which are not taken into account when calculating the spatial aliasing and spurious harmonics errors, will be negligible compared to that of orders lower than $\tilde{N}$.  
In particular, two additional orders were added to the last equation, compared to the criterion developed in \cite{rafaely2005analysis};
this was intended to ensure that non-modelled spatial-aliasing and spurious-harmonics errors will be significantly lower than $-40\,$dB, a value which will later be used for determing the variance of model-mismatch errors.
$\bm n_L(k)$ and $\bm n_M(k)$ are generated as in Eqs.~\eqref{eq:48} and \eqref{eq:50}, respectively. 
For this it is necessary to initially generate zero-mean normally-distributed i.~i.~d.~errors for constructing $\tilde{\bm n}_L(k)$ and $\tilde{\bm n}_M(k)$ from Eqs.~\eqref{eq:26} and \eqref{eq:28}, respectively. 
The variance of the errors in $\tilde{\bm n}_L(k)$ is set constant over frequency, and $40\,$dB lower than the average power of the elements in $\tilde{\bm h}_L(k)$ from Eq.~\eqref{eq:5} evaluated at $1\,$kHz.
Similarly for the SMA, the variance of the errors in $\tilde{\bm n}_M(k)$ is set constant over frequency, and $40\,$dB lower than the average power of the elements in $\tilde{\bm h}_M(k)$ from Eq.~\eqref{eq:10} evaluated at $1\,$kHz.  
Given the array transfer and error vectors, the total system error, the SLA error, and the SMA error are calculated as in Eqs.~\eqref{eq:51}, \eqref{eq:60}, and \eqref{eq:61}, respectively, and the SLA and SMA error bounds are calculated as in Eqs.~\eqref{eq:55}-\eqref{eq:58}. 
Finally, for smooth errors and error bounds, $\delta(k)$, $\delta_L(k)$, $\delta_M(k)$, $\mathit{m}_{L}(k)$, and $\mathit{m}_{M}(k)$ are calculated for thirty different realisations of $\tilde{\bm n}_L(k)$ and $\tilde{\bm n}_M(k)$, and then averaged.    

\begin{figure}[t]
\centering
\includegraphics[width = 1\linewidth]{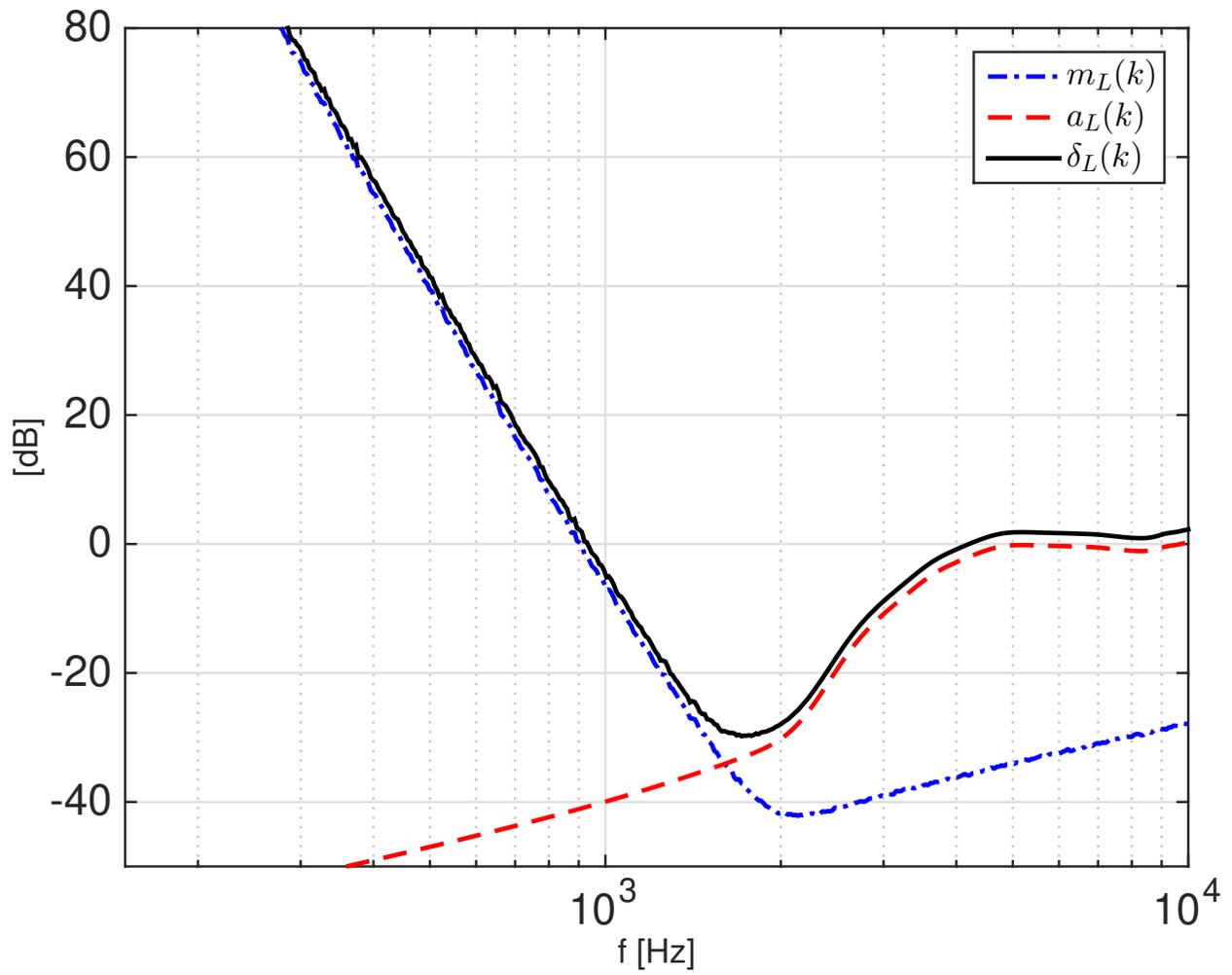}
\caption[(Color online) SLA errors and error bounds for \textit{SYS}1.]{(Color online) SLA errors and error bounds for \textit{SYS}1.}
\label{fig:jointdesign1a}
\end{figure}

\begin{figure}[t]
\centering
\includegraphics[width = 1\linewidth]{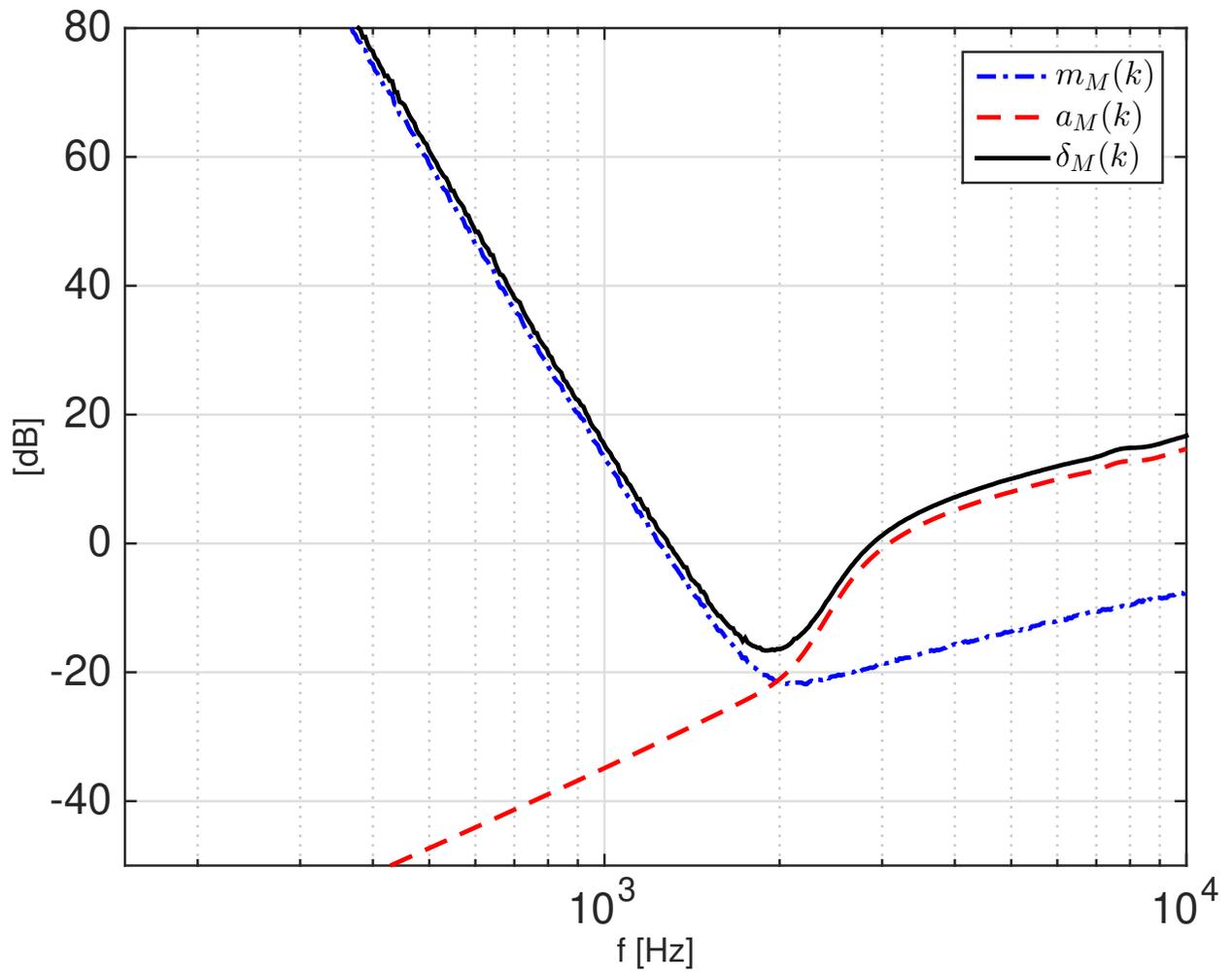}
\caption[(Color online) SMA errors and error bounds for \textit{SYS}1.]{(Color online) SMA errors and error bounds for \textit{SYS}1.}
\label{fig:jointdesign1b}
\end{figure}

Errors and error bounds are studied for the arrays of a MIMO system, and the SLA and SMA OFRs are defined. 
Figs.~\ref{fig:jointdesign1a} and \ref{fig:jointdesign1b} present the errors error bounds of the SLA and SMA, respectively, that comprise \textit{SYS}1.
For both arrays, the model-mismatch error bounds are dominant at low frequencies due to normalization, as explained in sec.~\ref{sec:normalization}.   
At high frequencies, the spurious-harmonics and spatial aliasing error bounds are seen to be dominant in Figs.~\ref{fig:jointdesign1a} and \ref{fig:jointdesign1b}, respectively. 
The total SLA error, $\delta_L(k)$, is also plotted on Fig.~\ref{fig:jointdesign1a} with a 2$\,$dB vertical offset added to this error for convenience of illustration, in order to avoid the overlap between the curves in the figure. 
$\delta_L(k)$ converges with the SLA model-mismatch error bound at low frequencies, and with the spurious-harmonics error bound at high frequencies. 
This shows that $\delta_L(k)$ is bounded by the superposition of $\mathit{m}_{L}(k)$ and $\mathit{a}_{L}(k)$, as suggested by Eq.~\eqref{eq:60}.
Similarly for the SMA, $\delta_M(k)$ is plotted in Fig.~\ref{fig:jointdesign1b} with a 2$\,$dB vertical offset, showing that it is bounded by the sum of errors as in Eq.~\eqref{eq:62}.  
$\delta_L(k)$ and $\delta_M(k)$ can be used for defining OFRs for the SLA and the SMA, respectively, and for a chosen threshold of tolerated error, denoted by $\sigma$. 
These OFRs are defined as:
\begin{eqnarray}\label{eq:65}
\mathit{O}_L & = & \{ k: \delta_L(k) \leq \sigma \} \\
&  \text{and} & \nonumber \\ 
\mathit{O}_M & = & \{ k: \delta_M(k) \leq \sigma \}, \label{eq:66}
\end{eqnarray}
for the SLA and SMA, respectively. 
For example, with $\sigma= 0\,$dB, the SLA and SMA OFRs are calculated to be $900\,$Hz-$5\,$kHz and $1.2\,$kHz$\,-3\,$kHz, respectively.
Moreover, the low- and high- frequency bounds of $\mathit{O}_L$ can be calculated as the frequencies at which $\mathit{m}_{L}(k) = \sigma $ and $\mathit{a}_{L}(k) = \sigma $, respectively, since $\delta_L(k)$ is bounded by $\mathit{m}_{L}(k)$ at low frequencies and $\mathit{a}_{L}(k)$ at high frequencies, as explained earlier. 
Similarly, the low- and high- frequency bounds of $\mathit{O}_M$ can be calculated using $\mathit{m}_{M}(k)$ and $\mathit{a}_{M}(k)$. 

\begin{figure}[t]
\centering
\includegraphics[width = 1.0\linewidth]{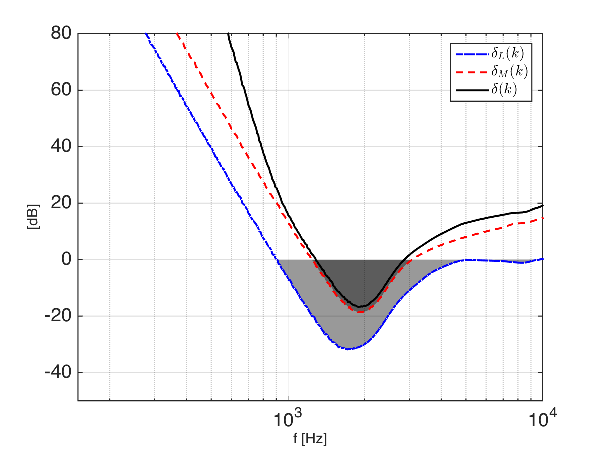}
\caption[(Color online) SLA, SMA, and system total error for \textit{SYS}1. $\mathit{O}_L$ and $\mathit{O}_M$, the SLA and SMA OFRs, respectively, calculated using $\sigma=0$, are indicated by the shaded areas.]{(Color online) SLA, SMA, and system total error for \textit{SYS}1. $\mathit{O}_L$ and $\mathit{O}_M$, the SLA and SMA OFRs, respectively, calculated using $\sigma=0$, are indicated by the shaded areas.}
\label{fig:jointdesign1c}
\end{figure}

\begin{figure}[t]
\centering
\includegraphics[width = 1\linewidth]{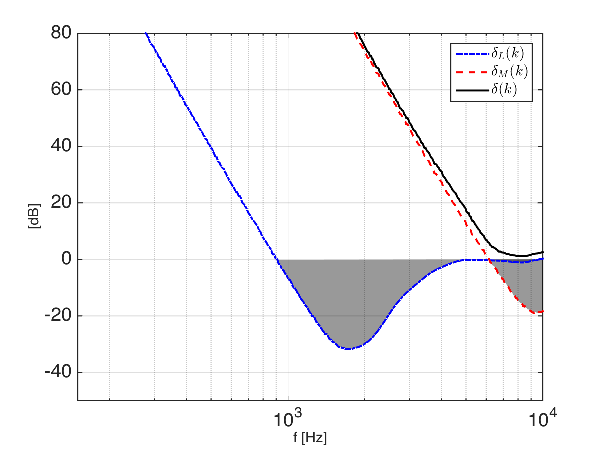}
\caption[Same as Fig.~\ref{fig:jointdesign1c} but for \textit{SYS}2.]{Same as Fig.~\ref{fig:jointdesign1c} but for \textit{SYS}2.}
\label{fig:jointdesign1d}
\end{figure}

The total error of a MIMO system is studied, and a system OFR is defined. 
Figs.~\ref{fig:jointdesign1c} and \ref{fig:jointdesign1d} present the SLA, SMA, and total system errors for \textit{SYS}1 and \textit{SYS}2, respectively. 
In particular, a 2$\,$dB vertical offset is added to the total system error, $\delta(k)$, in both figures for convenience of illustration. 
For \textit{SYS}1, $\delta(k)$ converges with the SMA total error when $\delta_M(k) \leq 0$. 
The latter shows that:
(i) it is sufficient to bound the SLA and SMA errors in order to ensure bounded energy of the total system error, as implied in Eq.~\eqref{eq:54new}; and
(ii) when the total error of the SMA is more significant than that of the SLA, or vice versa, $\delta(k)$ can be bounded by the dominant array error, which is $\delta_M(k)$ in this example.   
In Fig.~\ref{fig:jointdesign1d}, the total system error of \textit{SYS}2 exhibits values higher than $0\,$dB at all frequencies since at least one error, of either the SLA or the SMA, is greater than or equal to zero at all simulated frequencies.
This is a consequence of the smaller radius of the SMA comprising \textit{SYS}2, which results in a shift of the SMA error bounds to higher frequencies compared to those of \textit{SYS}1. 
For each system, an OFR, $\mathit{O}$, can be defined using $\delta(k)$ as 
\begin{eqnarray}\label{eq:64}
\mathit{O} & = & \{ k: \delta(k) \leq \sigma \}.     
\end{eqnarray}
With $\sigma= 0\,$dB, the OFR of \textit{SYS}1 is calculated to be $1.2\,$kHz -$3\,$kHz, and an empty set is calculated for \textit{SYS}2.
Since $\delta(k)$ is dominated by a single array error for different frequencies, $\mathit{O}$ can also be calculated as the intersection between $\mathit{O}_L$ and $\mathit{O}_M$ for thresholds $\sigma\leq 0\,$dB as: 
\begin{eqnarray}\label{eq:67}
\mathit{O} & = & \mathit{O}_L \cap \mathit{O}_M.   
\end{eqnarray}
This is also demonstrated in Figs.~\ref{fig:jointdesign1c} and \ref{fig:jointdesign1d};
the SLA and SMA OFRs, calculated using $\sigma = 0\,$dB, are indicated by the shaded areas. 
Their intersection, which is visibly darker in the figure, indicates the system OFR.    

The fact that the system OFR is defined by the intersection of the SLA and SMA OFRs motivates the definition of a matched MIMO system. 
A matched MIMO system is defined as a system for which the OFR of one array is a subset of the OFR of the other array. 
The system OFR in this case, $\mathit{\hat{O}}$, maintains the following condition: 
\begin{eqnarray}\label{eq:68}
\mathit{\hat{O}}& = &
\left\{ (   {{O}_L}, {{O}_M}):   \,\, {{O}_L}  \subseteq \mathit{O}_M \, \, \text{or}\,\,  {{O}_M}  \subseteq \mathit{O}_L     \right\}.
\end{eqnarray}
\textit{SYS}1 is therefore considered a matched system, as Eq.~\eqref{eq:68} is maintained in this case.  

The matching of the SLA and SMA in a MIMO system may not be unique. 
If the OFR of one array is a proper subset of the other arrays' OFR, i.e., $\mathit{O}_L \subset \mathit{O}_M$ or $\mathit{O}_M \subset \mathit{O}_L$, then there are multiple ways to match the arrays' OFRs and to maintain Eq.~\eqref{eq:68}. 
A practical criterion is proposed for matching the arrays' OFRs. 
Selecting the parameters of the system such that: 
\begin{eqnarray}\label{eq:70}
r_M N_L & = &  r_L N_M,
\end{eqnarray}
aligns the spurious-harmonics and spatial-aliasing error bounds of the SLA and SMA, respectively \cite{rafaely2005analysis}. 
This is because the frequency that maintains $kr_L=N_L$ and $kr_M=N_M$, which is considered as the upper operating frequency limit in the literature \cite{rafaely2005analysis}, is the same for both arrays if Eq.~\eqref{eq:70} is maintained.
This criterion is proposed since it is independent of the power of model-mismatch errors at the arrays, which may vary for different conditions and systems (such as the use of different equipment, for example). 
Note that Eq.~\eqref{eq:70} is maintained for \textit{SYS}1, while it is not maintained for \textit{SYS}2. 
As described earlier, the matching of a system may not be unique. 
Therefore, not maintaining Eq.~\eqref{eq:70} does not necessarily imply that a system is not matched. 
\medskip

\section{Extension of the system OFR}\label{sec:restrictedspatialresolution}
\justify
If the OFR of the MIMO system is not sufficiently large it can be extended by employing a lower SH order at one of the arrays.
Employing a lower SH orders leads to lowered spatial resolution of the array, on one hand;
however, it also leads to lower levels of the array errors and error bounds, thus increasing the width of the array's OFR and, potentially, the width of the system OFR. 

A criterion for selecting a lower SH order is developed, based on the criterion from Eq.~\eqref{eq:70} for matching a system. 
However, given that the arrays' radii are fixed, and that $N_L$ and $N_M$ are integers, it may not be possible to maintain this condition exactly.  
Therefore, a SH order is computed for minimizing $|r_M N_L -  r_L N_M|$, which aligns the spurious-harmonics and spatial-aliasing error bounds \cite{rafaely2005analysis}. 
In particular, since SH orders higher than $N_L$ and $N_M$ cannot be employed at the SLA and SMA, respectively, a SH order
\begin{eqnarray}\label{eq:71}
\check{N}_L & = & \underset{\check{N}_L\leq N_L}{\text{arg min}}  \, |r_M \check{N}_L -  r_L N_M|
\end{eqnarray}
is employed at the SLA if $r_M N_L >  r_L N_M$, or, if $r_M N_L <  r_L N_M$, a SH order
\begin{eqnarray}\label{eq:72}
\check{N}_M & = & \underset{\check{N}_M\leq N_M}{\text{arg min}}  \, |r_M N_L -  r_L \check{N}_M|
\end{eqnarray}
is employed at the SMA. 
For example, Eq.~\eqref{eq:72} gives $\check{N}_M = 2$ for \textit{SYS}2.

Errors and error bounds are formulated for \textit{SYS}2, employing $\check{N}_M = 2$.
Initially the SMA transfer vectors are calculated. 
$\bm \psi_M$ are generated, as in Eq.~\eqref{eq:19}, but using $\check{N}_M = 2$, instead of $N_M$. 
Generating $\hat{\bm \psi}_M$ from Eq.~\eqref{eq:46} requires the calculation of $\bm n_M(k)$ and $\bm z_M(k)$, using the same $\tilde{N}$ as in the previous section.   
$\bm n_M(k)$ requires the calculation of a modified $\bm \alpha_M$ (cf.~Eq.~\eqref{eq:50}).
A modified matrix, $\check{\bm \alpha}_M$, is defined up to the new SH order by truncating the first $(\check{N}_M+1)^2$ rows of $\bm \alpha_M$, from the previous section. 
$\bm z_M(k)$ requires the calculation of a modified $\bm \epsilon_M$ (cf.~Eq.~\eqref{eq:49}). 
A modified matrix, $\check{\bm \epsilon}_M$, is defined similarly to Eq.~\eqref{eq:31}, but using $\check{\bm \alpha}_M$, and is given by using the following equation:  
\begin{eqnarray}\label{eq:73}
\check{\bm \alpha}_M {\bm Y}_M    & = & [\check{\bm I}_M \,\,\check{\bm \epsilon}_M], 
\end{eqnarray}
where $\check{\bm I}_M$ is an $(\check{N}_M+1)^2 \times (\check{N}_M+1)^2$ identity matrix. 
Note that other methods can be considered for defining modified sampling weights, and employing such methods may alter the formulated SMA errors. 
These may include numerical designs suitable for $\check{N}_M $, and methods for aliasing cancellation \cite{rafaely2007spatial,alon2012spherical}, and are out of the scope of this work. 
Given the modified matrices, the SMA error and the total system error were calculated as in Eqs.~\eqref{eq:61} and \eqref{eq:51}, respectively, and the SMA error bounds were calculated as in Eqs.~\eqref{eq:57} and \eqref{eq:58}. 
Moreover, for smooth errors and error bounds, $\delta(k)$, $\delta_M(k)$, and $\mathit{m}_{M}(k)$ were averaged over thirty different realisations of $\tilde{\bm n}_L(k)$ and $\tilde{\bm n}_M(k)$, as in the previous section.    

\begin{figure}[t]
\centering
\includegraphics[width = 1\linewidth]{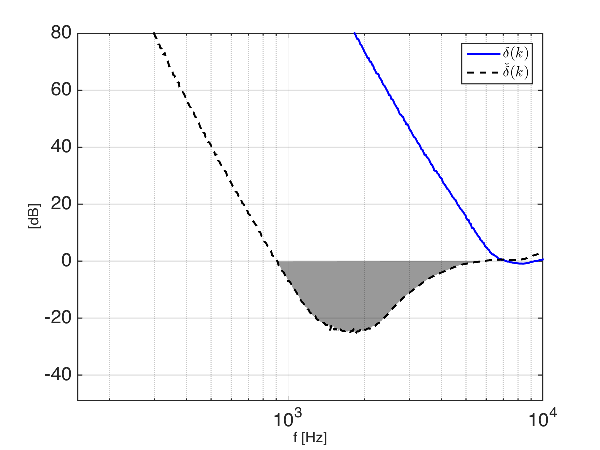}
\caption[(Color online) System errors for \textit{SYS}2: ${\delta}(k)$, before system matching (using $N_M$), and $\check{\delta}(k)$, after system matching (using $\check{N}_M = 2$). The modified OFR, calculated with $\sigma = 0$ after the matching of the system, is indicated by the shaded area.]{(Color online) System errors for \textit{SYS}2: ${\delta}(k)$, before system matching (using $N_M$), and $\check{\delta}(k)$, after system matching (using $\check{N}_M = 2$). The modified OFR, calculated with $\sigma = 0$ after the matching of the system, is indicated by the shaded area.}
\label{fig:extension1}
\end{figure}

The errors of \textit{SYS}2 are analyzed and its OFR is calculated for SH order $\check{N}_M = 2$ at the SMA. 
Fig.~\ref{fig:extension1} presents the total system error for the modified SH order, denoted $\check{\delta}(k)$, and for comparison, ${\delta}(k)$ of the original system from Fig.~\ref{fig:jointdesign1d} is also presented. 
Employing order $\check{N}_M$ is seen to result in a lower level of the total error bound, $\check{\delta}(k)$, compared to ${\delta}(k)$. 
In particular, using $\sigma= 0\,$dB, a modified OFR is calculated at $900\,$Hz-$5.6\,$kHz, demonstrating the effect of employing a lower SH order at one of the arrays on the system OFR. 
Finally, note that a SH order higher than $\check{N}_M$ can also be selected and, while employing a higher order may lead to an unmatched system, it is expected to improve the SMA spatial resolution.
\medskip

\section{Beamforming examples}\label{sec:desexamples}
\justify
The importance of MIMO system design and the implications of system matching on system performance and robustness against errors are studied in this section in the context of an analysis of directional RIRs. 
Beamforming is applied to the arrays in order to produce directional RIRs, with the aim of filtering a single reflection from the early refelction segment of an RIR. 
Positioning errors and transducer noise are also simulated in order to study the robustness against errors.  

\subsection{Setup}\label{sec:setup}
\begin{figure}[t]
\centering
\includegraphics[width = 1\linewidth]{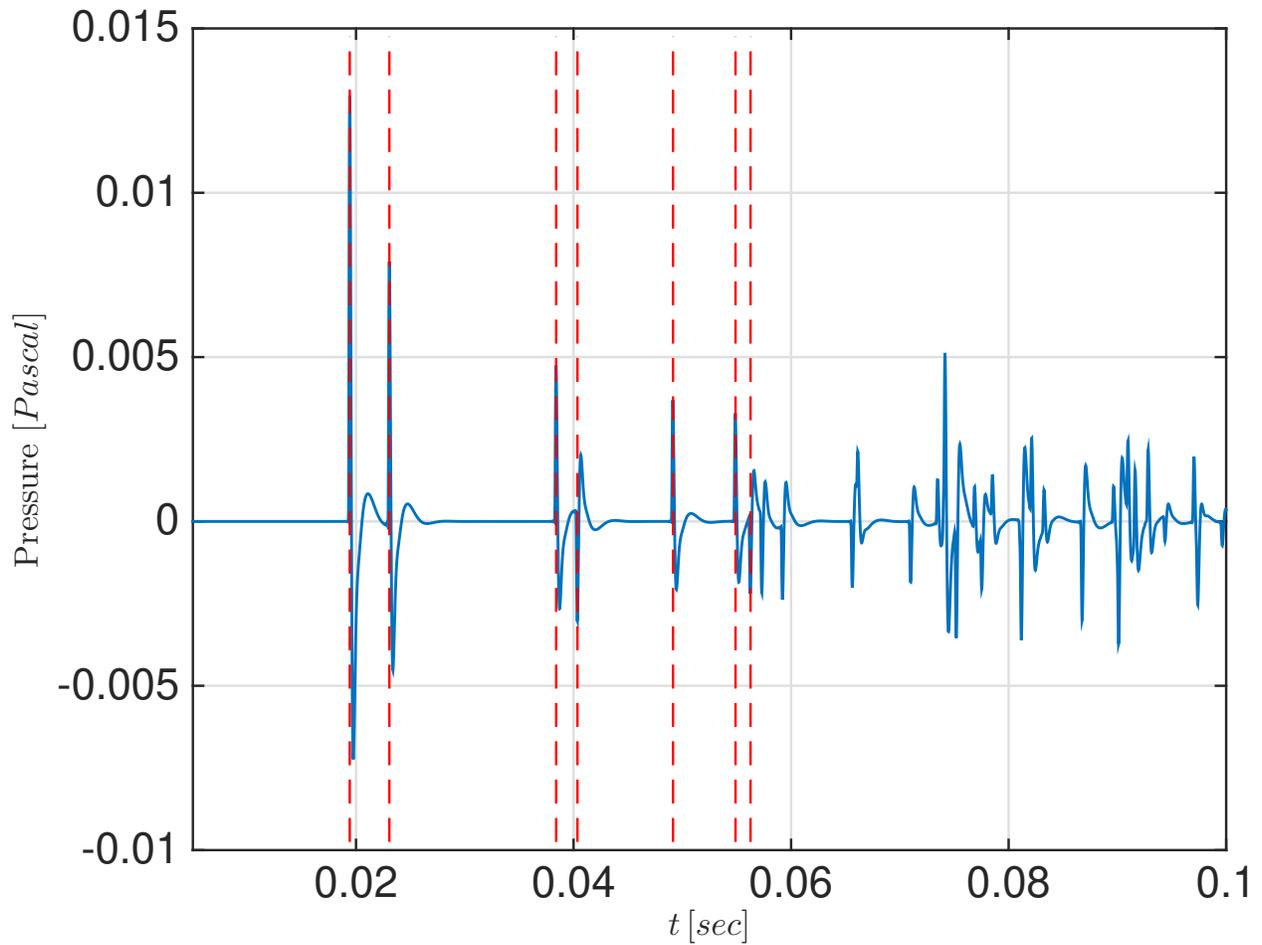}
\caption[(Color online) RIR simulated for omnidirectional loudspeaker and microphone. Vertical red-dashed lines indicate the time delays of direct sound and the first six reflections, listed in Table \ref{table:desexample1}.]{(Color online) RIR simulated for omnidirectional loudspeaker and microphone. Vertical red-dashed lines indicate the time delays of direct sound and the first six reflections, listed in Table \ref{table:desexample1}.}
\label{fig:desexample1}
\end{figure}

\justify
An example analysis is provided for MIMO systems positioned in a room. 
As in sec.~\ref{sec:jointdesign}, two systems are studied, which vary in only one system parameter, which is the SMA radius. 
The first system, \textit{SYS}1, is comprised of an SLA and an SMA with equal radii of $r_L = r_M = 0.2\,$m. 
$S= 36$ loudspeaker units are distributed uniformly on the surface of the SLA, which yields an order of $N_L=4$. 
The size of the loudspeakers' cap size is set to correspond to a 3''-diameter loudspeaker.
As for the SMA, $R = 50$ microphones are distributed using a Gaussian distribution, which yields $N_M = 4$.
For the second system, \textit{SYS}2, all system parameters are the same as for \textit{SYS}1 except for the radius of the SMA, which is set to $r_M = 0.04\,$m.
Both systems are positioned in a room with dimensions of $(25,15,10)\,$m and a reverberation time of approximately $750\,$ms. 
For both systems, transfer functions were simulated using MCRoomSim software \cite{wabnitz2010room}, with a sampling frequency of $f_s = 48$kHz, and with the SMA and the SLA positioned at $(15,8,3)\,$m and $(10,4,1.5)\,$m, respectively. 
The transfer functions were arranged in matrix form, as in Eq.~\eqref{eq:1}. 
Instead of generating model mismatch errors using the general error model employed in sec.~\ref{sec:mismatch errors}, in this section positioning errors were simulated for both arrays directly. 
Positioning errors were generated using a uniform distribution on the range $[-1^{\circ},1^{\circ}]$, for modelling arrays with a rotating mechanism, which typically encounter higher positioning errors compared to fixed arrays. 
These errors were then added to the angles of the loudspeaker and microphone positions. 
In addition to the positioning errors, transducer noise was also simulated for both arrays, with errors generated using a zero-mean Gaussian distribution. 
At the SLA, errors were added to the input signals of the different loudspeakers, which are given by products of the SLA input signal, $s(k)$, and the corresponding beamforming coefficient. 
The variance of the errors was set equal over frequency and for all loudspeakers, to produce a 25 dB average SNR for the loudspeakers at frequency 1$\,$kHz. 
For a given loudspeaker, the SNR is defined as the power of the corresponding input signal divided by the variance of the simulated errors. 
In particular, the SNR is defined with respect to the input signals that include beamforming coefficients of the non-normalized system in the space domain. 
This is becasue sampling weights and normalization were applied with beamforming in a single step in the following sections. 
At the SMA, errors were added to the input signals of the microphones. 
The variance of the errors was set equal over frequency and for all microphones, to produce a 25 dB average SNR for the microphones at 1$\,$kHz.
Finally, all responses were band-pass filtered to a range of $300\,$Hz-$1.9\,$kHz, as the arrays are expected to have significant errors outside this frequency band. 
Using the method from sec.~\ref{sec:jointdesign}, the OFR was calculated to be $550\,$Hz-$1.7\,$kHz for \textit{SYS}1 using $\sigma = 0\,$dB, and the system is matched since it maintains Eq.~\eqref{eq:68}. 
For \textit{SYS}2 an empty set is calculated for the OFR using $\sigma$. 
The same band-pass filter was applied to the responses of both systems so as to match the systems' temporal resolution.
For reference, the first $100\,$ms of the normalized RIR between an omnidirectional loudspeaker and an omnidirectional microphone is presented in Fig.~\ref{fig:desexample1}.  
This RIR is obtained by taking the element of the first row and first column in $\bm \Psi(k)$, and then employing an inverse FFT.
A typical RIR is composed of a sum of reflections for the direct sound and for sound reflected off the room walls. 
Each reflection is associated with a DOR around the SLA, a DOA around the SMA, and a time-delay (TD). 
For the simulated setup, the DORs, DOAs, and TDs of the direct sound and the first six reflections were calculated analytically and are summarized in Table \ref{table:desexample1}. 
The TDs are also plotted on Fig.~\ref{fig:desexample1} as vertical dashed lines. 

\subsection{Maximum DI beamformer}\label{sec:maxDI}
\justify
The first pair of directional RIRs is synthesised for the maximum DI beamformer \cite{van2004detection}.   
This beamformer is designed to maximize an array's DI, and since this index depends on the SH order employed at an array (and not on the array's radius), the same DI is calculated for the SLAs and the SMAs in both systems. 
Since the DI is a measure for the directivity of an array, the performance of both systems is considered to be equal in this example. 
Therefore, a study of directional RIRs synthesized using this beamfomer serves to compare the robustness against errors of the two systems. 

\begin{figure}[t]
\centering
\includegraphics[width = 1\linewidth]{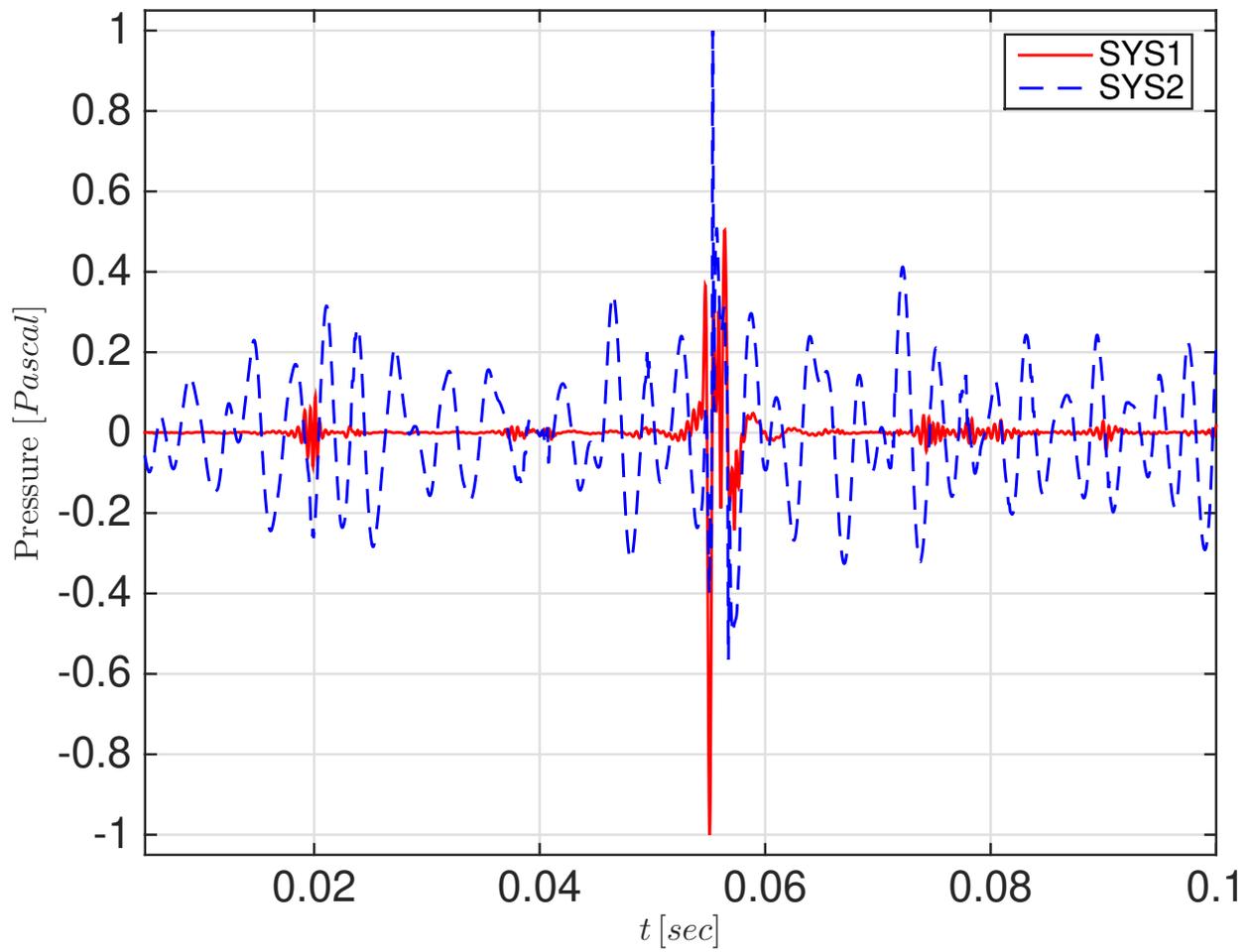}
\caption[(Color online) Directional RIRs synthesized by employing the maximum DI beamformer to the arrays of both systems.]{(Color online) Directional RIRs synthesized by employing the maximum DI beamformer to the arrays of both systems.}
\label{fig:desexample4}
\end{figure}

\begin{figure}[t]
\centering
\includegraphics[width = 1\linewidth]{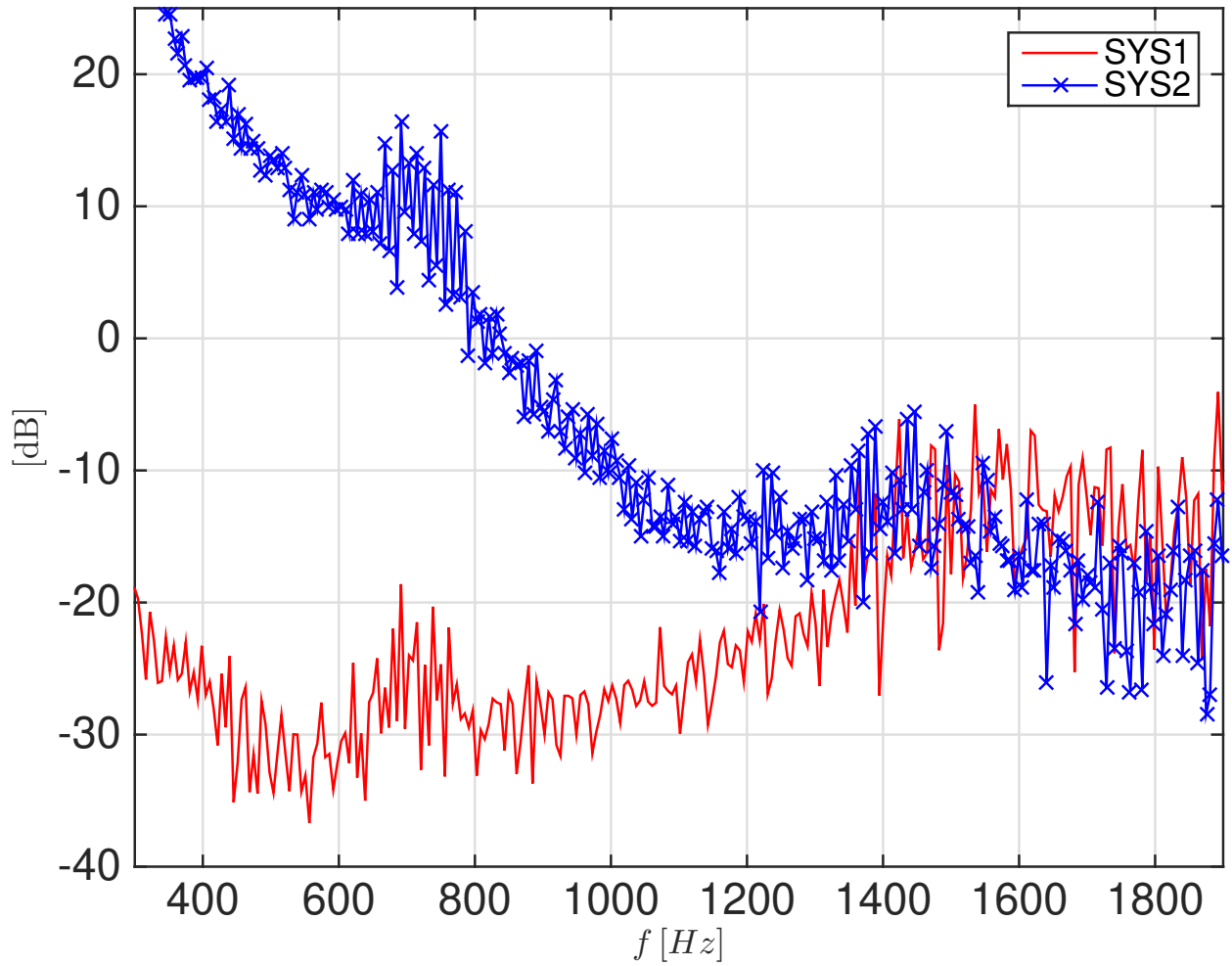}
\caption[(Color online) Total system error when the maximum DI beamformers were applied to the arrays of both systems.]{(Color online) Total system error when the maximum DI beamformers were applied to the arrays of both systems.}
\label{fig:desexample4a}
\end{figure}

Directional RIRs are generated, analyzed, and compared for both systems. 
These responses were synthesized using normalized beamforming vectors
\begin{eqnarray}\label{eq:75}
\bm \gamma(k)  & = &    \bm y_{N_L}(\bm \eta_5)\,\, \text{and} \\
\bm \lambda(k) &  = &  \bm y_{N_M}(\bm \beta_5) \label{eq:76}
\end{eqnarray} 
for the SLA and SMA, respectively. 
These vectors were preprocessed to account for sampling and normalization, and were then applied to the simulated system matrix, as in Eq.~\eqref{eq:1}. 
An inverse FFT was then employed for the resulting system outputs.
In particular, look directions $\bm \eta_5$ and $\bm \beta_5$ were selected, which correspond to the DOR and DOA of the fifth reflection from Table \ref{table:desexample1}.
Fig.~\ref{fig:desexample4} presents the directional RIRs for \textit{SYS}1 and \textit{SYS}2.
For ease of comparison, each response in this figure is normalized by its maximum absolute value.
A dominant peak  is evident at the TD of the fifth reflection for \textit{SYS}1. 
For \textit{SYS}2, on the other hand, energy is distributed for all times in the response, implying that a significant error is introduced to the system output in this case. 
A system error is defined, similarly to in Eq.~\eqref{eq:51}, but including beamforming at the arrays, as: 
\begin{eqnarray}\label{eq:77}
\Upsilon(k) & = & \frac{| \bm \lambda^\mathrm{H}(k) \bm \Psi(k) \bm \gamma(k)- \bm \lambda^\mathrm{H}(k) \hat{\bm \Psi}(k) \bm \gamma(k)   |}{ | \bm \lambda^\mathrm{H}(k) \bm \Psi(k) \bm \gamma(k)  | }. 
\end{eqnarray}
Fig.~\ref{fig:desexample4a} presents $\Upsilon(k)$ for both systems.
For smooth curves, $\Upsilon(k)$ was averaged over thirty realizations of positioining errors and transducer noise for both errors. 
It is evident that for \textit{SYS}1, $\Upsilon(k)$ is lower than $0\,$dB for low frequencies (within the frequency range defined by the band-pass filter). 
\textit{SYS}2, however, encounters errors higher than $0\,$dB at low frequencies, which explains the low-frequency error component in Fig.~\ref{fig:desexample4}. 
In summary, this beamforming example demonstrated that a matched system is more robust to errors, compared to an unmatched system, when the directivity (or performance) of the arrays in both systems, is equal.  

\subsection{Maximum white-noise gain beamformer}\label{sec:maxWNG}
\justify
The second pair of directional RIRs is synthesised for the maximum white-noise gain (WNG) beamformer \cite{van2004detection}.  
This beamformer is designed to maximize an array's WNG. 
Since the WNG depends on the number of elements in an array (and not on the array's radius) \cite{rafaely2005analysis,rafaely2011optimal}, the same value is calculated for the SLAs and the SMAs in both systems. 
Since the WNG is considered to be a measure for an array's robustenss against errors, system robustness to errors is considered to be equal for both arrays in this example. 
Therefore, a study of directional RIRs synthesized using this beamfomer serves to compare the performance of the two systems. 

\begin{figure}[t]
\centering
\includegraphics[width = 1\linewidth]{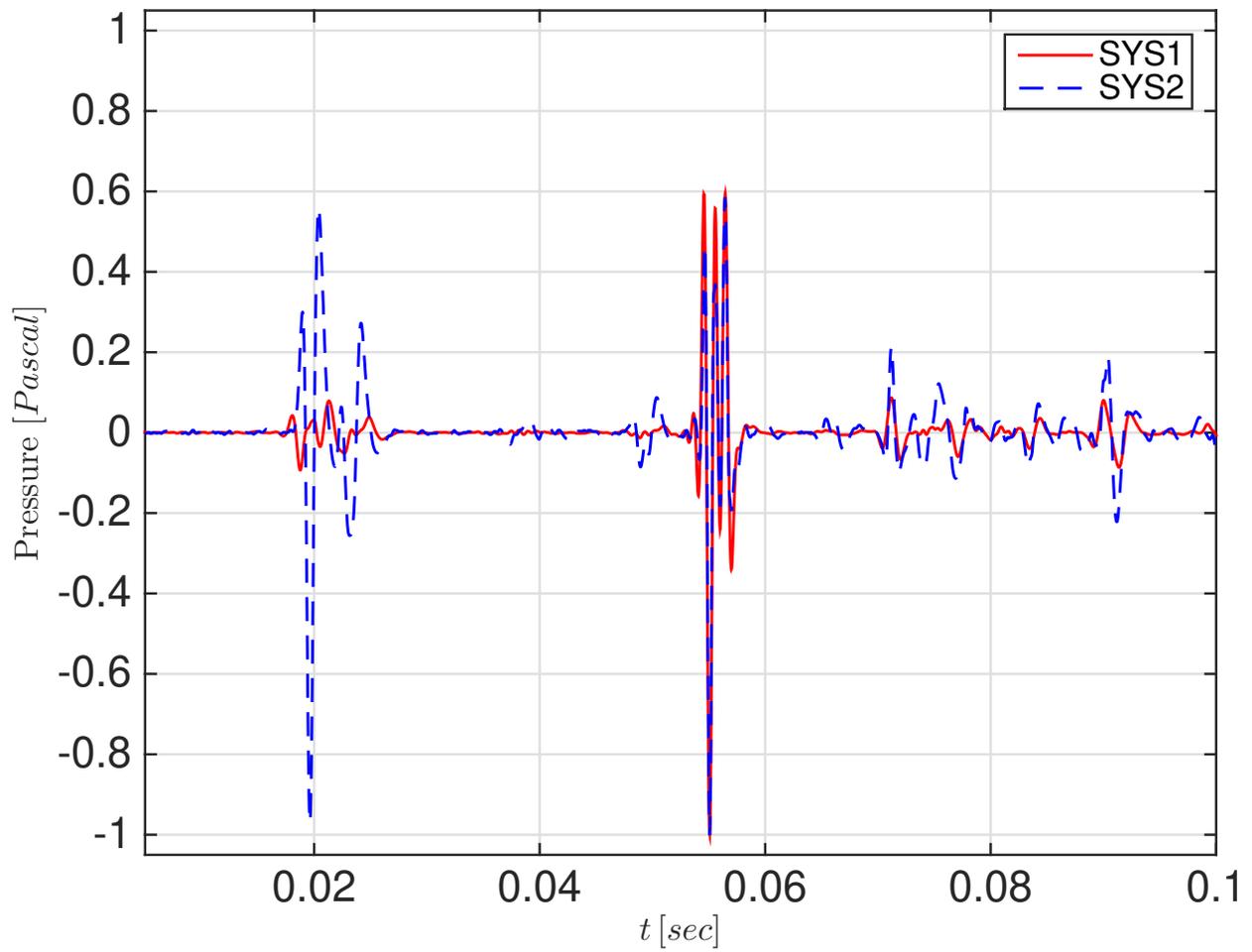}
\caption[(Color online) Directional RIRs synthesized by employing the maximum WNG beamformer to the arrays of both systems.]{(Color online) Directional RIRs synthesized by employing the maximum WNG beamformer to the arrays of both systems.}
\label{fig:desexample6}
\end{figure}

\begin{figure}[t]
\centering
\includegraphics[width = 1\linewidth]{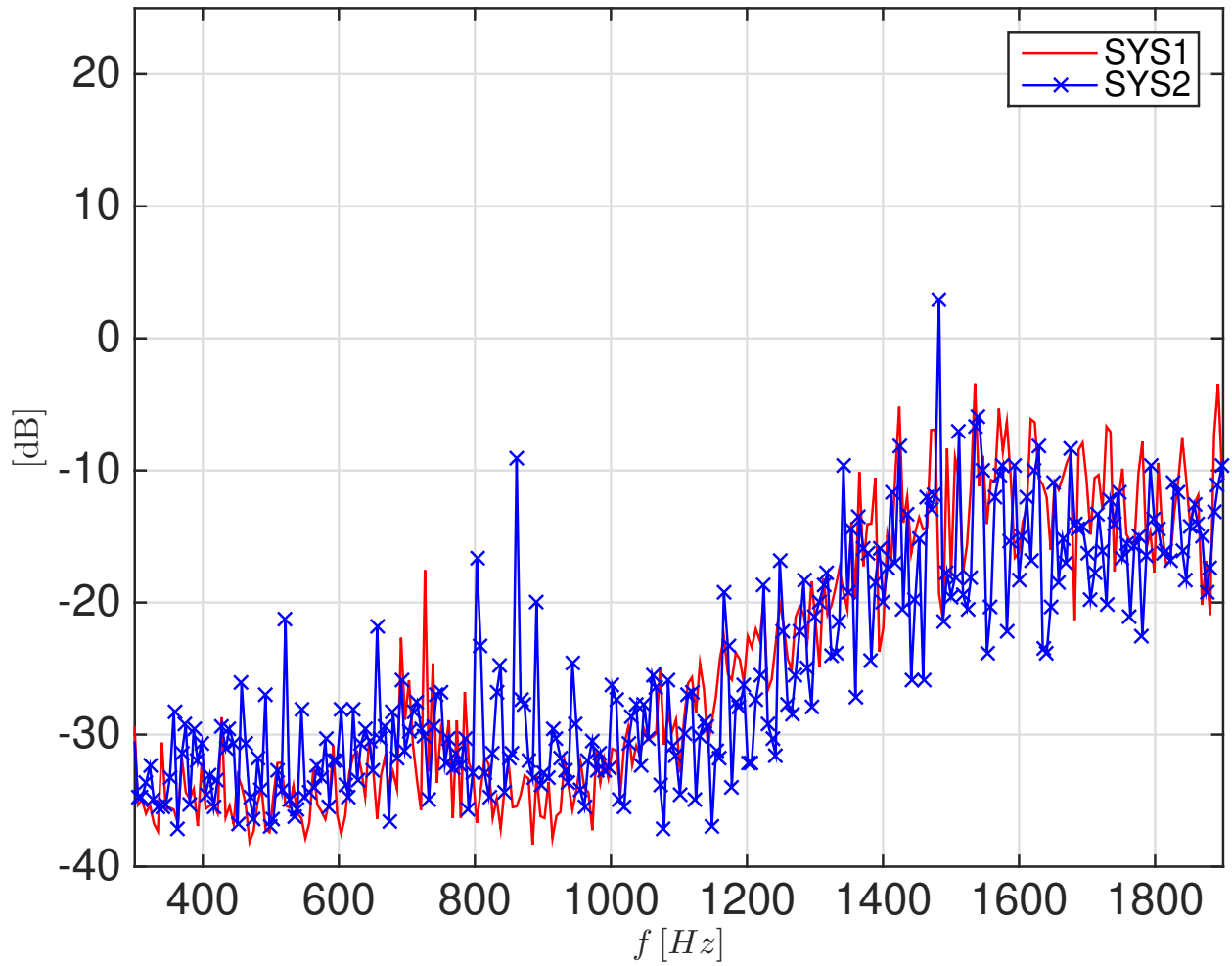}
\caption[(Color online) Total system error when maximum WNG beamformers were applied to the arrays of both systems.]{(Color online) Total system error when maximum WNG beamformers were applied to the arrays of both systems.}
\label{fig:desexamplehai}
\end{figure}

\begin{figure}[t]
\centering
\includegraphics[width = 1\linewidth]{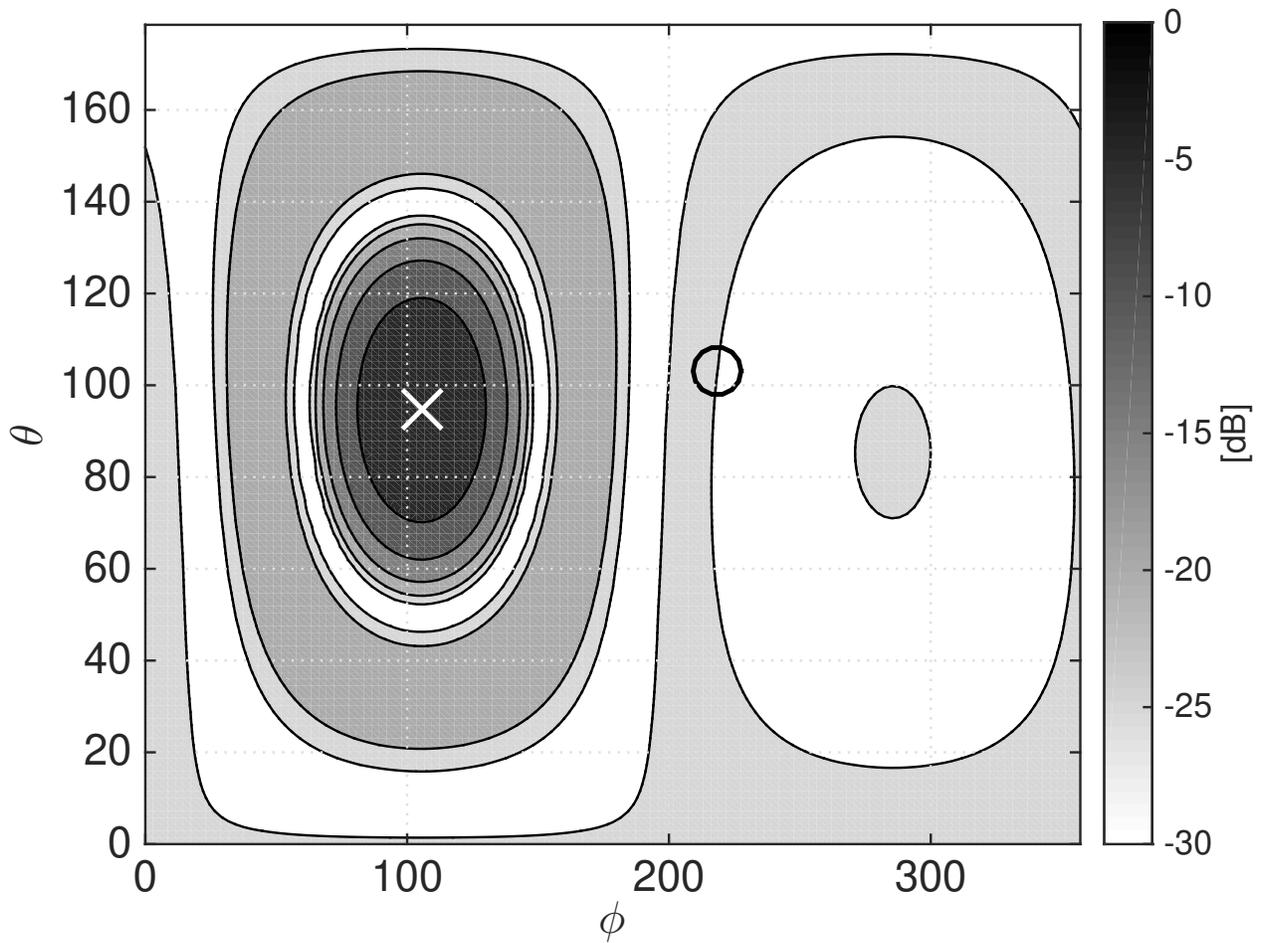}
\caption[SMA beampattern corresponding to beamforming vector $\bm u$ from Eq.~\eqref{eq:79} for \textit{SYS}1, calculated at a frequency of $1.1\,$kHz. `X'- and `O'-marks indicate the DOA of the fifth reflection and of the direct sound, respectively.]{SMA beampattern corresponding to beamforming vector $\bm u$ from Eq.~\eqref{eq:79} for \textit{SYS}1, calculated at a frequency of $1.1\,$kHz. `X'- and `O'-marks indicate the DOA of the fifth reflection and of the direct sound, respectively.}
\label{fig:desexample8}
\end{figure}

\begin{figure}[t]
\centering
\includegraphics[width = 1\linewidth]{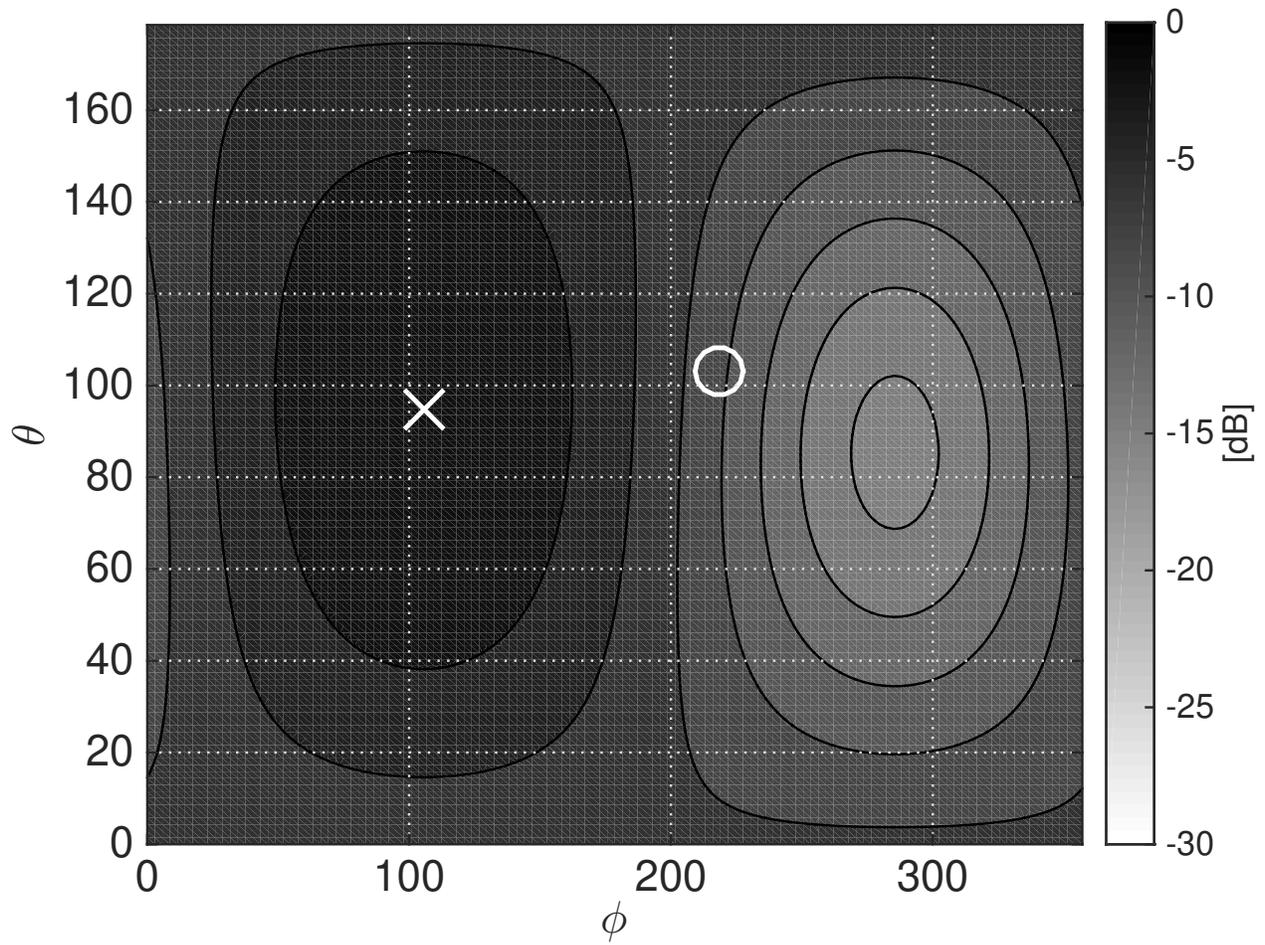}
\caption[Same as Fig.~\ref{fig:desexample8} but for \textit{SYS}2.]{Same as Fig.~\ref{fig:desexample8} but for \textit{SYS}2.}
\label{fig:desexample9}
\end{figure}

Directional RIRs are generated, analyzed, and compared for both systems. 
These responses were synthesized using normalized beamforming vectors
\begin{eqnarray}\label{eq:78}
\bm \gamma(k) & = &  \frac{   \bm G^\mathrm{H}(k) \bm G(k) \bm y_{N_L}(\bm \eta_5)}{  \sum_{n=0}^{N_L} \frac{2n+1}{4\pi}|g_{n}(kr_L)|^2  }\,\, 
\text{and} \\
\bm \lambda(k)  & = &  \frac{  \bm B(k) \bm B^\mathrm{H}(k) \bm y_{N_M}(\bm \beta_5)}{  \sum_{n=0}^{N_M} \frac{2n+1}{4\pi}|b_{n}(kr_M)|^2  }\label{eq:79}
\end{eqnarray} 
for the SLA and SMA, respectively. 
As in the previous example, the beamfroming coefficient vectors were preprocessed to account for sampling and normalization, and were then applied to the simulated system matrix, as in Eq.~\eqref{eq:1}, and an inverse FFT was employed for the resulting system outputs.
Fig.~\ref{fig:desexample6} presents the directional RIRs for \textit{SYS}1 and \textit{SYS}2.
In this case, no error component is evident in the directional RIRs of both systems. 
This can be explained by the similar errors $\Upsilon(k)$ for the systems, as seen in Fig.~\ref{fig:desexamplehai}. 
Moreover, only one significant peak is evident for \textit{SYS}1 in Fig.~\ref{fig:desexample6}, while an additional peak is seen at the TD of the direct sound in the response of \textit{SYS}2. 
This can be explained by the SMA directivity of \textit{SYS}1, which is higher than that of \textit{SYS}2, as further illustrated below. 
To evaluate the directivity of the SMAs, the SMA beam pattern is plotted for \textit{SYS}1 and \textit{SYS}2 in Figs.~\ref{fig:desexample8} and \ref{fig:desexample9} for $1.1\,$kHz. 
For ease of comparison, the beampatterns power at the fifth DOA, i.e., $\bm \beta_5$, is normalized to $0\,$dB for both figures, and a dynamic range of $30\,$dB is used. 
Moreover, on the figures, `X'- and `O'-marks are plotted to indicate the DOA of the fifth reflection and that of the direct sound, respectively. 
Comparing the beampattern of the two systems shows that \textit{SYS}1 is more directive than \textit{SYS}2 at the evaluated frequency. 
Moreover, at the DOA of the direct sound, the power of the beampattern is calculated at $-25.11\,$dB and $-7.63\,$dB for \textit{SYS}1 and \textit{SYS}2, respectively, explaining the significant difference in the attenuation of the direct sound illustrated in Fig.~\ref{fig:desexample6}. 

In conclusion, in this example it was shown that a matched system has improved performance, compared to an unmatched system, when considering directivity and robustness. 
\medskip

\section{Conclusions}\label{sec:conclusions}
\justify
A framework for designing an SLA and an SMA in a MIMO system was presented, based on the formulation of a system model that includes errors due to spatial sampling and model mismatch at both arrays. 
Using this framework, a matched system was defined, for which the OFRs of the SMA and the SLA are matched. 
The superiority of a matched system over an unmatched system was demonstrated in several simulation examples, in which spatial filtering was applied to the systems in order to synthesize directional RIRs.  
The development of improved methods for analysis and synthesis of sound fields using MIMO systems is proposed for future work.

\section{ACKNOWLEDGMENTS}\label{sec:ACK}
\justify
This research was supported by The Israel Science Foundation (Grant No.~146/13) and the Chateaubriand fellowship.

\newpage
\justify
\begin{table}
\centering
\begin{tabular}{| c | c | c | c|}
 \hline
reflection & TD [sec]  & DOR $(\theta, \phi)$ [deg] & DOA $( \xi, \omega)$ [deg]  \\ \hline \hline
Direct sound & 0.0192 & (76.82, 38.66) & (103.18, 218.66) \\ \hline
1 & 0.0228 & (125.10, 38.66) & (125.10, 218.66) \\ \hline
2 & 0.0382 & (83.42, 292.62) & (96.58, 247.38) \\ \hline
3 & 0.0401 & (109.09, 292.62) & (109.09, 247.38) \\ \hline
4 & 0.0489 & (22.45, 38.66) & (22.45, 218.66) \\ \hline
5 & 0.0546 & (85.41, 74.48) & (94.59, 105.52) \\ \hline
6 & 0.0560 & (103.54, 74.48) & (103.54, 105.52) \\ \hline
\end{tabular}
\caption{TDs, DORs, and DOAs of the direct sound and the first six reflections in the normalized RIR, presented in Fig.~\ref{fig:desexample1}.}
\label{table:desexample1}
\end{table}


\newpage

\clearpage

\clearpage

\clearpage

\end{document}